\documentclass[twocolumn]{aastex63}

\usepackage{graphicx}
\usepackage{longtable}
\usepackage{amsmath}
\usepackage{booktabs}
\usepackage{multirow}
\usepackage{float}
\usepackage{lineno}

\received{January 4, 2021}
\revised{January 22, 2021}
\accepted{February 16, 2021}

\submitjournal{ApJ}

\shorttitle{AGN Outflows in Dwarf Galaxies}
\shortauthors{Bohn et al.}

\graphicspath{{./}{figures/}}

\begin{document}

\title{Near-Infrared Coronal Line Observations of Dwarf Galaxies hosting AGN-driven Outflows}

\email{tbohn002@ucr.edu}

\author[0000-0002-4375-254X]{Thomas Bohn}
\affil{University of California, Riverside, Department of Physics \& Astronomy \\
900 University Ave., Riverside, CA 92521}

\author[0000-0003-4693-6157]{Gabriela Canalizo}
\affiliation{University of California, Riverside, Department of Physics \& Astronomy \\
900 University Ave., Riverside, CA 92521}

\author[0000-0002-3158-6820]{Sylvain Veilleux} 
\affiliation{Department of Astronomy, University of Maryland, College Park, MD 20742, USA}
\affiliation{Joint Space-Science Institute, University of Maryland, College Park, MD 20742, USA}

\author[0000-0003-3762-7344]{Weizhe Liu}
\affiliation{Department of Astronomy, University of Maryland, College Park, MD 20742, USA}

\begin{abstract}

We have obtained Keck NIR spectroscopy of a sample of nine M$_\star$ $<$ 10$^{10}$ M$_\odot$ dwarf galaxies to confirm AGN activity and the presence of galaxy-wide, AGN-driven outflows through coronal line (CL) emission. We find strong CL detections in 5/9 galaxies (55$\%$) with line ratios incompatible with shocks, confirming the presence of AGN in these galaxies. Similar CL detection rates are found in larger samples of more massive galaxies hosting type 1 and 2 AGN. We investigate the connection between the CLs and galaxy-wide outflows by analyzing the kinematics of the CL region, as well as the scaling of gas velocity with ionization potential of different CLs. In addition, using complementary Keck KCWI observations of these objects, we find that the outflow velocities measured in [\ion{Si}{6}] are generally faster than those seen in [\ion{O}{3}]. The galaxies with the fastest outflows seen in [\ion{O}{3}] also have the highest [\ion{Si}{6}] luminosity. The lack of $J$-band CN absorption features, which are often associated with younger stellar populations, provides further evidence that these outflows are driven by AGN in low mass galaxies.

\end{abstract}


\section{Introduction} \label{sec:intro}

It is now well accepted that supermassive black holes (SMBHs) lie at the center of most massive galaxies. Strong correlations have supported the idea that they evolve with their host galaxies and feedback from the AGN has been shown to regulate star formation \citep[for a review, see][]{Kormendy2013_ARAA}. Powerful, AGN-driven outflows are believed to suppress star formation \citep{Rupke2017,U2019} in high mass galaxies, leading their hosts to the well-defined red sequence.

The general consensus regarding feedback in dwarf galaxies (M$_\star$ $<$ 10$^{10}$ M$_\odot$) is that stellar processes, such as starbursts and supernovae, provide the main source of quenching \citep[eg.,][]{Veilleux2005,Veilleux2020,Heckman2017}. However, the growing rate of AGN detections in dwarf galaxies \citep{Reines2013,Moran2014,Sartori2015} necessitates a closer look at AGN feedback in the low-mass regime. Evidence of AGN-driven feedback in dwarf galaxies is already starting to emerge. \citet{Penny2018} present a sample of dwarf galaxies with AGN line ratios and kinematically disturbed gas at their centers, possibly due to AGN feedback. In addition, \citet{Bradford2018} found a sample of isolated ($>$ 1Mpc), gas-depleted dwarfs with optical line ratios indicative of AGN activity, which suggests that AGN may play a role in clearing gas from dwarf galaxies. 

The shallower gravitational potential wells of dwarf galaxies allow feedback mechanisms to have a more profound effect on the global interstellar medium (ISM). Although this means dwarf galaxies are more susceptible to external environmental effects, strong outflows may be capable of driving gas beyond the dark matter halo. To investigate this possibility, \citet{Manzano2019}, hereafter MK19, observed a sample of 29 isolated dwarf galaxies with optical and IR signatures suggestive of AGN activity. Nine of these show fast outflows with median speed of $\sim$645 km s$^{-1}$, higher than the median host escape velocities of $\sim$300 km s$^{-1}$. If the outflows are of AGN origin, this would suggest that AGN feedback could eject material beyond the dark matter halo and have a substantial impact on the star formation rate. Furthermore, follow-up integral field spectroscopy done by \citet[][hereafter L20]{Liu2020} of this sample show spatially extended outflows up to 3 kpc in a number of these targets. In addition, they detect outflow velocities greater than 500 km s$^{-1}$ in six of these galaxies. Indeed, a small but non-negligible fraction (up to 6$\%$) of the ionized outflowing gas has the necessary speeds to escape their galaxy.

However, line ratios from star-forming processes can sometimes mimic those of AGN and some contamination exists between the AGN and star-forming regions of the Baldwin, Phillips $\&$ Terlevich \citep[hereafter BPT;][]{Baldwin1981} diagram \citep[eg.,][]{Allen2008,Rich2011}. Moreover, L20 found core-collapse supernovae to be energetic enough to drive the mass outflow rates ($3\times10^{-3}$ --- 0.3 $\rm{M_{\odot}\; yr^{-1}}$) seen in their sample and thus the contribution of stellar processes in driving the outflows can not be formally ruled out.

To further characterize the kinematics of outflows, coronal lines (CLs) have been used as an additional tracer of outflows. CLs are forbidden transitions from highly ionized ($>$100 eV) species with widths suggesting the coronal line region lies between the broad and narrow line regions \citep{DeRobertis1984,DeRobertis1986,Penston1984,Erkens1997,Rodriguez2002,Rodriguez2006}. Due to the high energies needed for their ionization, CLs are excellent indicators of AGN activity. They are often observed blueshifted relative to the systemic velocity of the host galaxy and thus are believed to be linked with outflows. \citet{Muller2011} have measured outflow speeds upwards of 1500 km s$^{-1}$ through CL emission and the velocity fields suggest the outflows are of biconical shape, with collimation likely due to the torus \citep[see Standard Model,][]{Antonucci1993}.

In this article, we present follow-up NIR spectroscopy of the nine dwarf galaxies from MK19 and L20 that show optical AGN line ratios and fast outflows. Through NIR diagnostics, we aim to confirm the presence of AGN activity and characterize the outflows through NIR emission lines. Details of the sample selection, observations, and data reduction are summarized in Section \ref{sec:Observations}. Analysis of the data, including spectra fitting and AGN diagnostics are covered in Section \ref{sec:Analysis}. In Section \ref{sec:Discussion}, we discuss the NIR emission lines and outflow characteristics. Throughout this article, we adopt a standard $\Lambda$CDM cosmology with $H_0$ = 70 km s$^{-1}$ Mpc$^{-1}$, $\Omega_M$ = 0.3, and $\Omega_\Lambda$ = 0.7.

\section{Data and Observations} \label{sec:Observations}

\subsection{Sample Selection} \label{Selection}

\begin{deluxetable*}{ccccccccc}
\caption{Observation Log} 
\label{tab:obs_log}
\tablehead{\colhead{Galaxy} & \colhead{Date} & \colhead{Redshift} &
\colhead{Exp. Time\tablenotemark{a}} & \colhead{Slit PA} & \colhead{Ext. Ap.} & \colhead{S/N\tablenotemark{b}} & \colhead{Airmass} & 
\colhead{Telluric} \\
\colhead{} & \colhead{(YYYY-mm-dd)} & \colhead{} & \colhead{} & \colhead{(degrees)} & \colhead{(arcsec)} & \colhead{} & \colhead{} & \colhead{}}
\startdata
SDSS J010005.93-011058.89\tablenotemark{c} & 2017-10-28 & 0.05151 & 8 x 240s & 51 & 1.89 & 25 & 1.12 & HD18571\\
SDSS J081145.29+232825.72 & 2019-01-24 & 0.01573 & 12 x 240s & 45 & 1.63 & 13 & 1.58 & BD-001836\\
SDSS J084025.54+181858.99 & 2019-01-24 & 0.01498 & 8 x 240s & 101 & 1.63 & 17 & 1.34 & HD74721\\
SDSS J084234.50+031930.68\tablenotemark{d} & 2019-01-24 & 0.02882 & 8 x 240s & 277 & 1.63 & 18 & 1.28 & HD74721\\
SDSS J090613.76+561015.22\tablenotemark{d} & 2019-01-24 & 0.04664 & 12 x 240s & 239 & 1.51 & 20 & 1.31 & HD92573\\
SDDS J095418.15+471725.11\tablenotemark{d} & 2019-01-24 & 0.03266 & 10 x 240s & 144 & 1.63 & 18 & 1.18 & HD92573\\
SDSS J100551.18+125740.65\tablenotemark{d} & 2018-10-24 & 0.00949 & 6 x 240s & 65 & 1.90 & 23 & 1.60 & HD77332\\
SDSS J100935.66+265648.99\tablenotemark{d} & 2019-01-24 & 0.01436 & 8 x 240s & 226 & 1.63 & 18 & 1.03 & HD86986\\
SDSS J144252.78+205451.67 & 2019-01-24 & 0.04262 & 9 x 240s & 101 & 1.52 & 14 & 1.24 & HD124773
\enddata
\tablenotetext{a}{Exposures were typically done in ABBA nodding.}
\tablenotetext{b}{Average continuum signal-to-noise ratio across all orders.}
\tablenotetext{c}{SDSS J0100-0110 was observed with NIRSPEC. All other targets were observed with NIRES.}
\tablenotetext{d}{Indicates galaxies with coronal line detections. See Section \ref{subsec:coronal_detections}.}
\end{deluxetable*}

Our sample of nine dwarf galaxies comes from the AGN sample of MK19. Briefly, \citet{Reines2013}, \citet{Moran2014}, and \citet{Sartori2015} have identified hundreds of dwarf galaxies with optical and IR signatures indicative of AGN activity. MK19 created a subsample of candidate AGN whose optical line ratios place them above the star-forming region of the BPT and Veilleux $\&$ Osterbrock 1987 \citep[hereafter VO87;][]{Veilleux1987} line ratio diagrams or that have \ion{He}{2} $\lambda$4686 emission \citep[see][]{Shirazi2012}. They obtained Keck LRIS \citep[eg.,][]{Oke1995,Rockosi2010} spectroscopy of 29 of these galaxies, nine of which show blue asymmetries in their [\ion{O}{3}] $\lambda$5007 profile. This blue wing is often regarded as an indicator of outflowing gas, where the asymmetry arises due to the redshifted gas being blocked from our line of sight. These nine dwarf galaxies with spatially extended outflows form our sample for this article. Their stellar masses range from 8.77 $<$ log($\rm{M_{*}/M_{\odot}}$) $<$ 9.97 (median of 9.34) and all have redshift z $<$ 0.05.

\subsection{Observations and Reductions} \label{Observe}

NIR spectroscopy was obtained on three separate dates: on 2017-10-29 using Keck II NIRSPEC \citep{McLean1998}, and on 2018-10-25 and 2019-01-25 with Keck II NIRES \citep{Wilson2004}. NIRSPEC is a NIR echelle spectrograph with a wavelength coverage from 0.9 --- 5.5 $\mu$m. The NIRSPEC-7 filter was used in low-resolution mode with a cross-dispersion angle of 35.31 degrees. This resulted in a wavelength coverage of $\sim$1.97 --- 2.39 $\mu$m. The $42''\times0.76''$ slit was used and a spectral resolution of 196 km s$^{-1}$ (R $\approx$ 1500) at 2.20 $\mu$m was measured with a seeing of $\sim$0.50$''$. Observations throughout the night were done under mostly clear conditions. Note that these observations were done before the NIRSPEC upgrade. NIRES is a NIR echelette spectrograph with the slit being 18$''\times0.55''$ and the wavelength coverage set from 0.94 --- 2.45 $\mu$m across five orders. There is a small gap in coverage between 1.85 and 1.88 $\mu$m, but this is a region of low atmospheric transmission. The average spectral resolution of the five orders range between 84 --- 89 km s$^{-1}$ (R $\approx$ 3400) and these differ less than 5$\%$ for each galaxy. Observations on 2018-10-25 were taken under variable and heavy cloud cover, however the majority of our sample was observed on 2019-01-25, where cloud cover was light. Individual exposures for all sets of observations were four minutes each and were done using the standard ABBA nodding. A telluric standard star, typically of A0 spectral class with measured magnitudes in $J$, $H$, and $K$-bands, was observed either directly before or after the target galaxy to correct for the atmospheric absorption features. Typical airmass differences with the target were below 0.10. A summary of the NIR observations is shown in Table \ref{tab:obs_log}.

In addition to the Keck NIRES and LRIS observations, follow-up optical IFU observations with Keck KCWI \citep{Morrissey2018} and Gemini GMOS \citep{Allington2002, Gimeno2016} were done to obtain high spatial resolution of the outflows. The details of this analysis are discussed in L20. 

Four of our targets (J0811+2328, J0906+5610, J0954+4717, and J1005+1257) were observed with the \textit{Chandra X-ray Observatory}. \citet{Baldassare2017a} report hard X-ray emission that is likely originating from the AGN in J0906+5610 and J0954+4717. Additionally, \citet{Wang2016} provide fluxes, corrected for galactic absorption, for J0811+2328 and J1005+1257, from which we calculated a luminosity using the cosmology listed above. We discuss these results in Section \ref{subsec:Detection_Rate}.

The data were reduced using two modified pipelines. The first provided flat fielding and a robust background subtraction by using techniques described in \citet{Kelson2003} and \citet{Becker2009}. In short, this routine maps the 2D science frame and models the sky background before rectification, thus reducing the possibility of artifacts appearing due to the binning of sharp features. The sky subtraction attained with this procedure is excellent, despite the strong OH lines present in the NIR; the procedure is also quite insensitive to cosmic rays and hot pixels, and is reliable regardless of skyline intensity.

Rectification, telluric correction, wavelength calibration, and extraction were all done with a slightly modified version of \textsc{REDSPEC}.\footnote{\url{https://www2.keck.hawaii.edu/inst/nirspec/redspec.html}} Telluric correction was done by dividing by the spectrum of the telluric standard star and multiplying by a blackbody curve of the same temperature. Strong OH skylines were used for wavelength calibration and the 1D spectra were then median combined. Flux calibration of individual exposures was done using the telluric star and the Spitzer Science center unit converter\footnote{http://ssc.spitzer.caltech.edu/warmmission/propkit/pet/magtojy/} to convert the magnitude of the star to the associated flux in that band. A small corrective factor ($<$5\%) was introduced due to the differences between the center NIR bands and that of the wavelength coverage.

\section{Analysis} \label{sec:Analysis}

\subsection{Spectral Fitting} \label{subsec:Fitting}

\subsubsection{NIR Fitting} \label{subsubsec:NIR_fit}

We fit all NIR spectra using \textsc{emcee}, an affine invariant Markov Chain Monte Carlo (MCMC) ensemble sampler \citep{Foreman-Mackey2013}. A narrow Gaussian component, along with a second order polynomial for the continuum, were fit simultaneously for each emission line. We determined whether a second Gaussian component was needed to fit the emission lines using the following \textit{F}-test: \textit{F} = $(\sigma_{single})^{2}/(\sigma_{double})^{2}$, where $\sigma$ is the standard deviation of the residuals using either single or double Gaussian components. If \textit{F} $>$ 2.0, then adding an extra component is justifiable and we added a component that is constrained to be broader and lower in amplitude than the first component to avoid degeneracies (see Appendix \ref{appendix:CL_detail} for example figures). Both the narrow and broad components were treated as Gaussians with amplitude, full-width half-maximum (FWHM), and velocity offset from rest-frame wavelength as free variables. We list all detected emission line fluxes and widths in Appendix \ref{appendix:CL_detail}. Note that we only report 2$\sigma$ detections and those with a FWHM greater than the resolution element.

As discussed in Section \ref{subsec:Detection_Rate}, we also use NIR absorption features in our analysis. These were also fit with MCMC in a similar fashion, the main difference being the amplitude was restricted to be negative. The fitting was done simultaneously with that of emission lines in order to keep the continuum level consistent across all measurements. Although we ran the \textit{F}-test as defined above, only one gaussian was needed for all the absorption fits. The widths and depth of the absorption features used in our analysis are also listed in Appendix \ref{appendix:CL_detail}.

\subsubsection{SDSS Fitting} \label{subsubsec:SDSS_fit}

SDSS spectra are available for our entire sample and these provide full wavelength coverage from 4000 \AA\; to 9000 \AA, which includes the [\ion{O}{2}] $\lambda\lambda$7320, 7330 doublet that is outside the LRIS coverage. The SDSS spectra were fit using Bayesian AGN Decomposition Analysis for SDSS Spectra \citep[{\textsc{BADASS,}\footnote{\url{https://github.com/remingtonsexton/BADASS2}}}][]{Sexton2020}, a spectral analysis tool that fits the stellar and Fe II features. Absorption features were fit using the penalized Pixel Fitting \citep[\textsc{pPXF}\footnote{\url{https://www-astro.physics.ox.ac.uk/~mxc/software/}};][]{Cappellari2004} and \ion{Fe}{2} emission was fit using \ion{Fe}{2} templates. All of these components were fit simultaneously, allowing for a detailed and robust analysis of the spectrum. 

The code allows the user to test for the presence of outflows by setting various constraints on parameters such as minimum amplitude, minimum width, and velocity offset. The profile of [\ion{O}{3}] $\lambda$5007 is often used as a tracer of outflows since it is isolated and in a region free of significant absorption. Through [\ion{O}{3}] $\lambda$5007, we detect strong outflows in the SDSS spectra of six galaxies: J0100-0110, J0811+2328, J0842+0319, J0906+5610, J0954+4717, and J1005+1257 at a $>$ 95$\%$ confidence based on the \textit{F}-test model comparison included in the code. J1009+2656 is a little more uncertain with a 89$\%$ confidence. We do not detect outflows (confidence $<$ 55$\%$) for J0840+1818 and J1442+2054, likely due to the lower resolution of SDSS. These results are consistent with the KCWI results from L20 who found outflows in all targets but J0840+1818 (note that J1442+2054 was not observed with KCWI).

\subsection{Extinction} \label{subsec:Extinction}

We note that L20 reports extinction values calculated from the H$\gamma$/H$\beta$ Balmer decrement. However, for J0100-0110 and J0811+2328, they used H$\alpha$/H$\beta$ measurements from SDSS due to weak H$\gamma$ emission in their spectra. In addition, they did not observe J1442+2054. We thus opted to use the H$\alpha$/H$\beta$ Balmer decrement measured from SDSS for our entire sample. 

To quantify the extinction, we used the intrinsic line ratio of H$\alpha$/H$\beta$ = 3.1, typically used for AGN \citep[eg.,][]{Veilleux1987,Osterbrock1992}, and a Cardelli reddening law \citep{Cardelli1989} with an extinction factor of $R_V$ = 3.1. We used the narrow-line flux measurements from the SDSS data (see Section \ref{subsubsec:SDSS_fit}), where we have decomposed the H$\alpha$ and H$\beta$ emission into narrow and broad (outflow) components. For the two galaxies, J0840+1818 and J1442+2054, where no outflow component was detected, we used the full emission line to obtain a flux. These values of the Balmer decrement and E(B-V) for each galaxy are listed in Table \ref{tab:extinction} and all flux values in Appendix \ref{appendix:CL_detail} reflect extinction corrected fluxes. Note that three galaxies, J0840+1818, J0842+0319, and J0906+5610, have Balmer decrements slightly below the intrinsic ratio so we did not apply any extinction correction to them.

Our results are slightly different but generally agree with those calculated by L20. The discrepancies are likely caused by the different line ratios (H$\gamma$/H$\beta$ versus H$\alpha$/H$\beta$) and line profiles used, where we only used the narrow profile while L20 used the full (narrow + broad) profile. When comparing the effect of these two methods, the difference in flux measurements amounts to $<$ 5\% for the majority of our sample. For the rest of this article, we use these extinction corrected flux values unless otherwise specified.

\begin{deluxetable}{ccc}
\caption{Measured Extinction Values of the Sample} 
\label{tab:extinction}
\tablehead{\colhead{Galaxy} & \colhead{H$\alpha/\rm{H}\beta$ Balmer Decrement} & \colhead{E(B-V)}}
\startdata
J0100--0110 & 3.47$\pm$0.27 & 0.115\\
J0811+2328 & 3.71$\pm$0.34 & 0.182\\
J0840+1818 & 3.06$\pm$0.17 & 0.000\\
J0842+0319 & 3.03$\pm$0.14 & 0.000\\
J0906+5610 & 2.94$\pm$0.18 & 0.000\\
J0954+4717 & 3.25$\pm$0.08 & 0.049\\
J1005+1257 & 5.75$\pm$0.27 & 0.624\\
J1009+2656 & 3.65$\pm$0.10 & 0.166\\
J1442+2054 & 3.69$\pm$0.15 & 0.176
\enddata
\tablecomments{An intrinsic ratio of H$\alpha/\rm{H}\beta$ = 3.1 and a Cardelli reddening law were used.}
\end{deluxetable}

\begin{figure*}
\centering
\epsscale{1.1}
\plotone{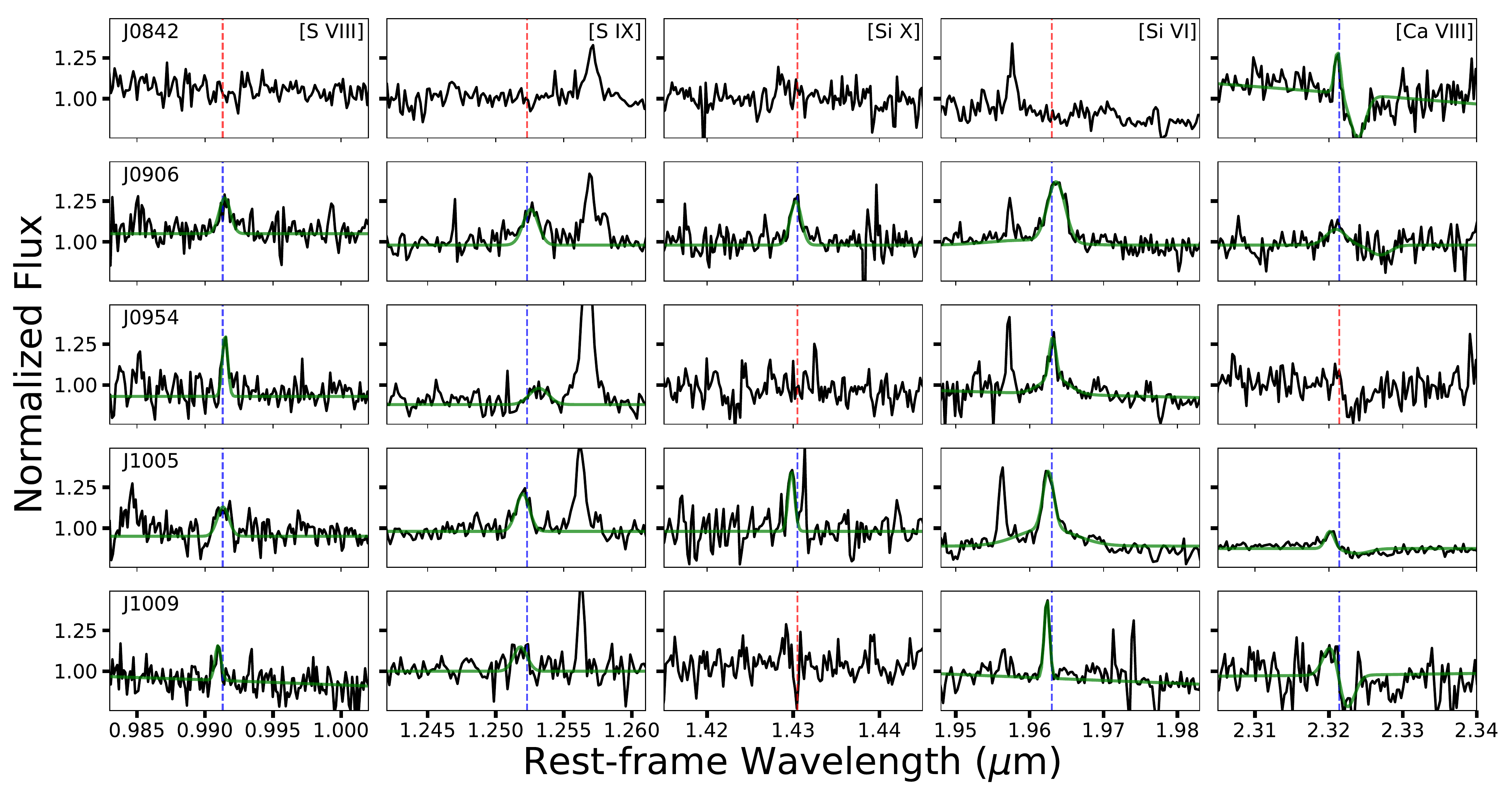}
\caption{Zoom-in plots of the spectral regions around the most prominent NIR CLs, where the flux has been normalized to unity and the systemic redshift was used to shift the spectra to rest-frame wavelength. Dotted blue (detections) and red (non-detections) indicate the rest-frame wavelength of CLs. MCMC fits to the emission and absorption line profiles are shown as solid green lines.\\ \label{fig:Main_CL_box}}
\end{figure*}

\subsection{Coronal Line Detections} \label{subsec:coronal_detections}

55$\%$ (5/9) of our sample have NIR CL emission within the spectral window of 0.94 --- 2.45 $\mu$m (see Table \ref{tab:obs_log}). For a similar spectral window and galaxy type, Sy1 and Sy2, this rate is consistent with others found in the literature: 66$\%$ \citep[36 out of 54,][]{Rodriguez2011}, 25$\%$\citep[5 out of 20,][]{Mason2015}, and 43$\%$ \citep[44 out of 102,][]{Lamperti2017}. Our detections predominately come from sulfur and silicon species: [\ion{S}{8}] 0.9913 $\mu$m, [\ion{S}{9}] 1.2523 $\mu$m, [\ion{Si}{10}] 1.4305 $\mu$m, and [\ion{Si}{6}] 1.9630 $\mu$m. We also detect [\ion{Ca}{8}] 2.3214 $\mu$m but it falls within the CO(3-1) absorption band at 2.3226 $\mu$m, making their measurements more uncertain. All of the detected NIR CLs are shown in Figure \ref{fig:Main_CL_box}, where we include our MCMC fits to the emission line profiles. We do not detect other common NIR CLs, such as [\ion{Fe}{13}] 1.0747 $\mu$m, [\ion{S}{11}] 1.9196 $\mu$m, and [\ion{Al}{9}] 2.0450 $\mu$m. The most prominent CL is [\ion{Si}{6}] and is detected in four of the five galaxies with CL detections. This is not surprising since it has a lower IP level than other CLs and is not near any absorption features. In three of these four galaxies, we find broad [\ion{Si}{6}] emission and thus use a two component fit to the profile, in accordance with the \textit{F}-test mentioned in Section \ref{subsubsec:NIR_fit}. Details of the multi-component fits to [\ion{Si}{6}] and plots of the entire spectra are presented in Appendix \ref{appendix:CL_detail}.

\begin{figure}
\centering
\epsscale{1.1}
\plotone{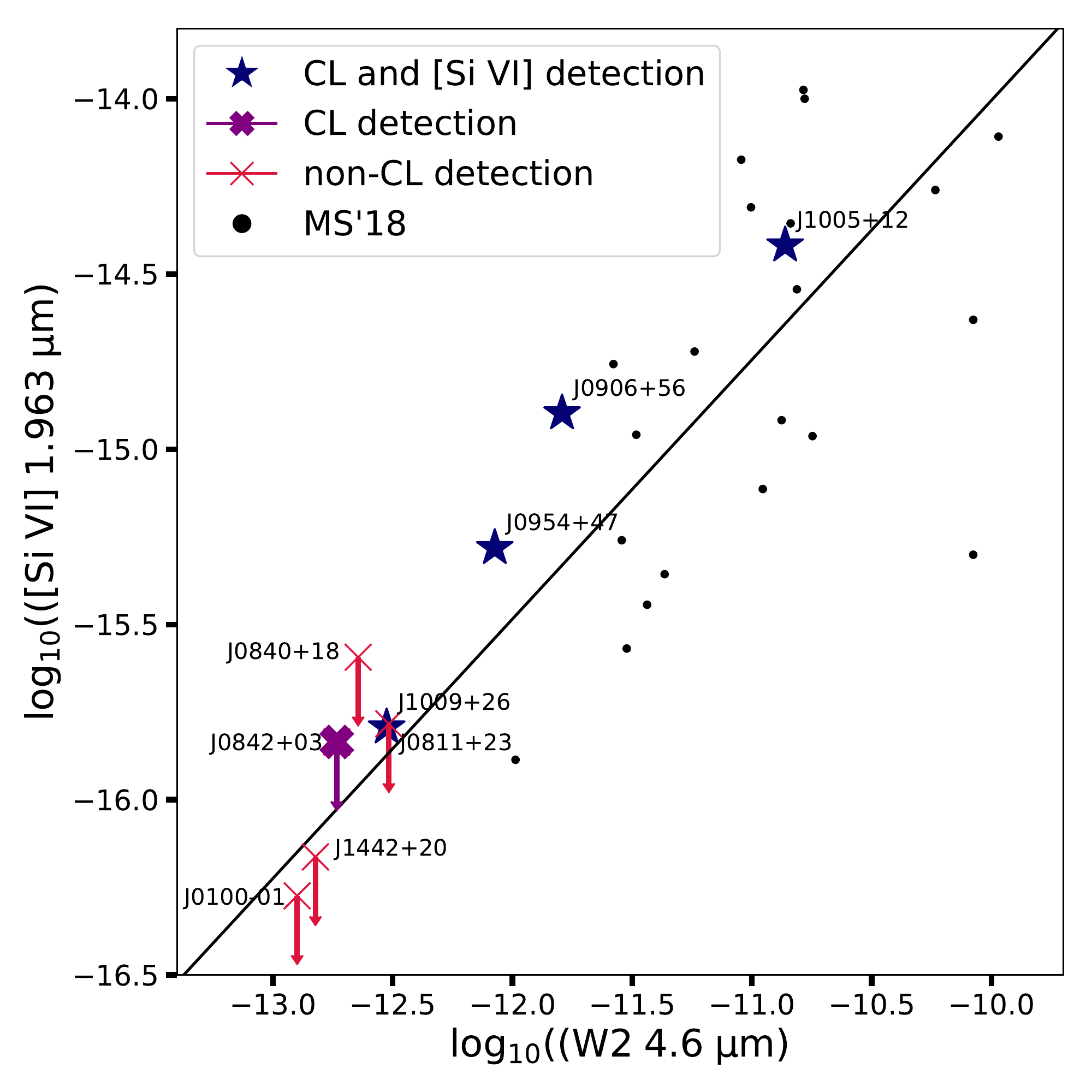}
\caption{[\ion{Si}{6}] flux vs W2 (4.6 $\mu$m) flux. Galaxies in our sample with detected [\ion{Si}{6}] emission are shown as blue stars. J0842+0319, the one object with a CL detection but no [\ion{Si}{6}] emission, is shown as a purple cross. Objects without any CL detections are shown as red crosses. For these latter two, the crosses represent 3$\sigma$ upper limit fluxes to [\ion{Si}{6}]. The solid line is the best-fit line to the data found in \citet{Muller2018}. 
\label{fig:SiVI_W2}}
\end{figure}

No NIR CL emission is detected in the other four galaxies. To investigate this, we plot total [\ion{Si}{6}] flux versus WISE \textit{W}2 (4.6 $\mu$m) flux in Figure \ref{fig:SiVI_W2}. Note that one galaxy, J0842+0319, has a CL detection, [\ion{Ca}{8}], but does not show any [\ion{Si}{6}] emission. Upper limit fluxes to the non-[\ion{Si}{6}] detections were calculated by integrating over a Gaussian with a width equaling the resolution element and amplitude equalling the 1$\sigma$ noise level where [\ion{Si}{6}] should appear. This value was multiplied by three to obtain the 3$\sigma$ upper limit that is plotted in Figure \ref{fig:SiVI_W2}. We also plot the AGN sample (black dots) from \citet{Muller2018}, from which we derive the plotted line of best fit. This relation (J. Cann, private communication) provides an expected value for [\ion{Si}{6}] based on \textit{W}2 flux. Our non-detections lie below most of our [\ion{Si}{6}] measurements, suggesting that deeper observations may be required to detect any CL emission. Further discussion on our non-detections is covered in Section \ref{subsec:Detection_Rate}.

The optical data also reveal a number of CLs, particularly in those with NIR CL detections. 55$\%$ (5/9) of our sample have strong optical CL emission. The most common detections in our sample include [\ion{Ne}{5}] $\lambda$3426, [\ion{Fe}{7}] $\lambda$6087, and [\ion{Fe}{10}] $\lambda$6374. Less common lines include [\ion{Ne}{5}] $\lambda$3346, and [\ion{Fe}{7}] $\lambda$5721. Most galaxies that have NIR CL emission also have optical CL emission, the only exceptions are J0840+1818 and J0842+0319, where the latter only shows NIR CL emission. For J0840+1818, MK19 report the detection of [\ion{Ne}{5}] $\lambda$3426 but we do not find any NIR CLs. No other significant detections are seen in the rest of the sample. Additional details for each galaxy are covered in Appendix \ref{appendix:CL_detail}. For the remainder of this article, we label galaxies as either having NIR CL emission or not.

\subsection{AGN Diagnostic Plots} \label{subsec:AGN_diagnostic}

\begin{figure*}
\gridline{\fig{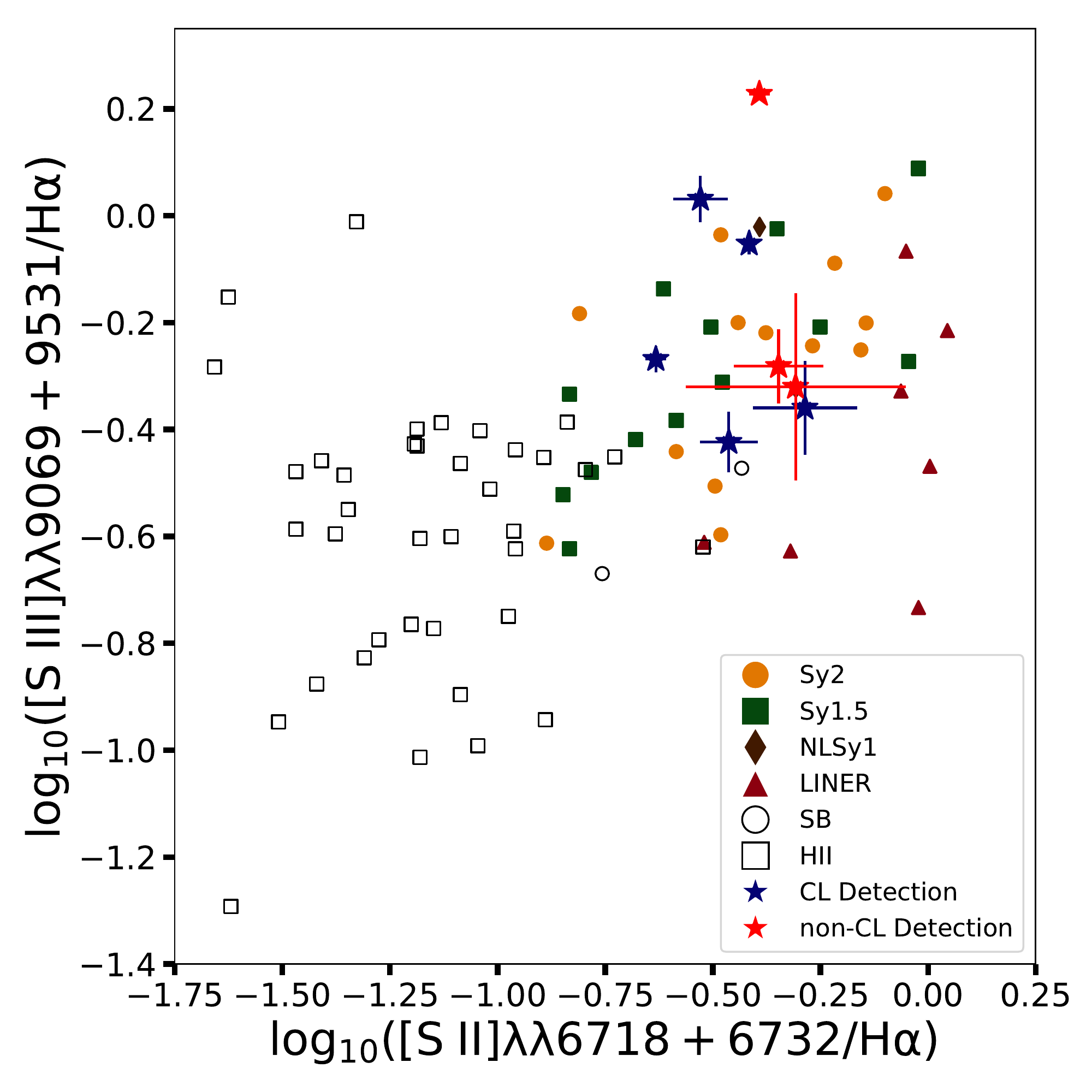}{0.34\textwidth}{}
\fig{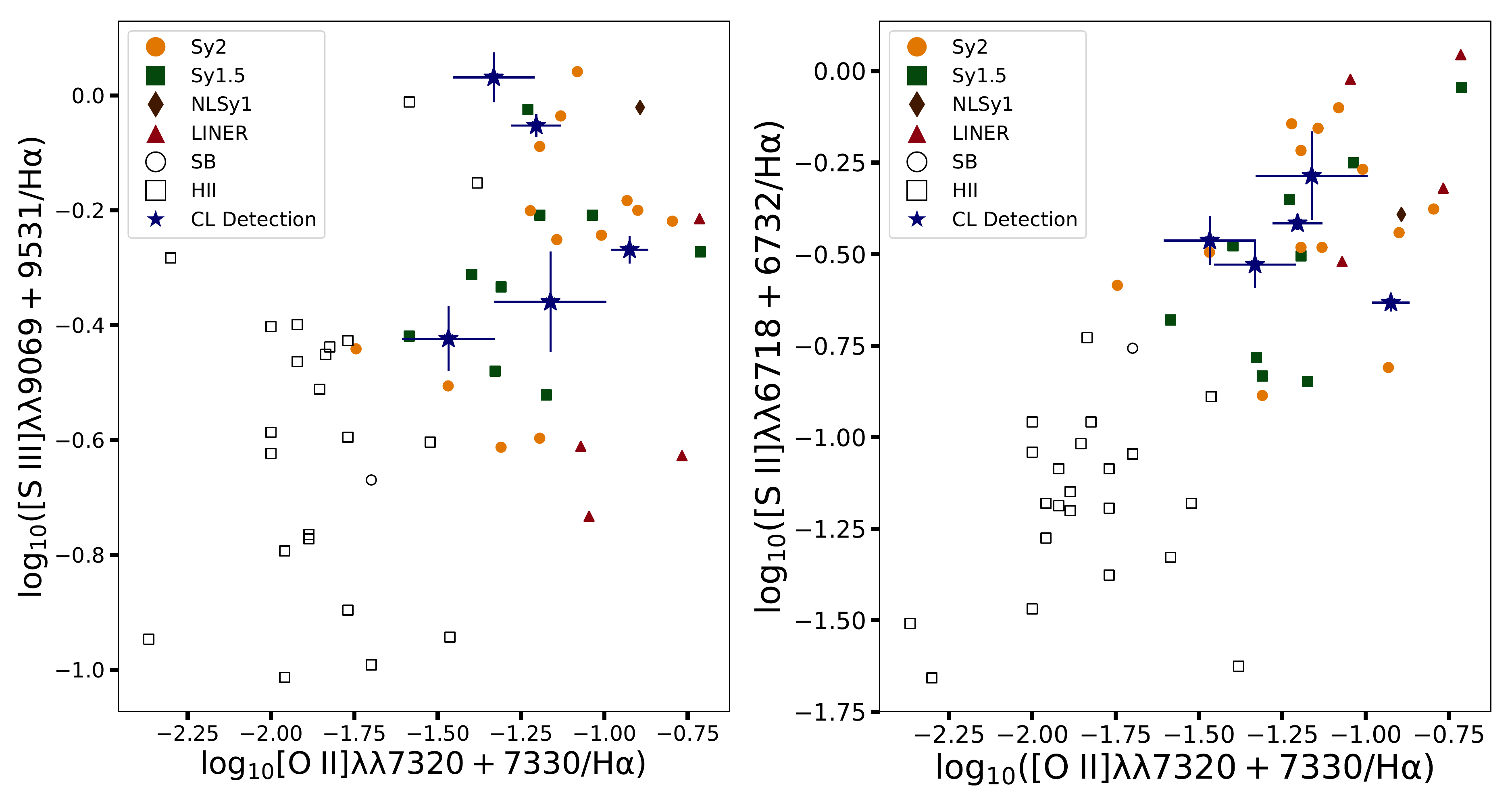}{0.64\textwidth}{}}
\caption{AGN diagnostic plots derived by \citet{Osterbrock1992}. We plot our sample as blue (CL detection) and red (non-CL detection) stars. Our sample falls within the scatter of other galaxies hosting AGN, further confirming the presence of AGN.
\label{fig:AGN_relations}}
\end{figure*}

Although there is evidence for AGN activity in our dwarf galaxy sample, including optical BPT/VO87 AGN line ratios, the addition of NIR lines allows us to run additional AGN diagnostics. With the inclusion of [\ion{S}{3}] $\lambda$9531, we can test our sample with the AGN diagnostics presented in \citet{Osterbrock1992}. Specifically we use the [\ion{S}{3}] $\lambda\lambda$9069+9531/H$\alpha$ versus [\ion{S}{2}] $\lambda\lambda$6718+6732/H$\alpha$, [\ion{S}{3}] $\lambda\lambda$9069+9531/H$\alpha$ versus [\ion{O}{2}] $\lambda\lambda$7320+7330/H$\alpha$, and [\ion{S}{2}] $\lambda\lambda$6718+6732/H$\alpha$ versus [\ion{O}{2}] $\lambda\lambda$7320+7330/H$\alpha$  (their Figures 4, 5 and 7). All three of these relations are plotted in Figure \ref{fig:AGN_relations}, where we include the AGN and star-forming samples from \citet{Osterbrock1992}. We overplot narrow-line fluxes of our sample as stars and exclude the galaxies with no measurable [\ion{O}{2}] $\lambda\lambda$7320, 7330. J0100-0110 is also omitted since its [\ion{S}{3}] $\lambda$9531 is outside the wavelength coverage of NIRSPEC. Since [\ion{S}{3}] $\lambda$9069 is outside the wavelength coverage of NIRES, we use the intrinsic flux ratio of [\ion{S}{3}] $\lambda$9531/$\lambda$9069 = 2.6 to obtain flux values for [\ion{S}{3}] $\lambda$9069.

In all three relations, our dwarf sample falls within the scatter of the AGN sample in \citet{Osterbrock1992}. As discussed in Section \ref{subsec:coronal_detections}, not every galaxy in our sample show CLs. To check if this has any effect, we separate our sample into two groups: those with CL emission and those without. We find that they all lie in similar regions and we see no clear distinction between them in all three relations. 
While no single line ratio can confidently separate AGN activity from star formation, our dwarf sample shows optical and NIR line ratios that, when considered collectively, strongly suggest the presence of AGN.

\section{Discussion} \label{sec:Discussion}

\subsection{Detection Rate of Coronal Lines} \label{subsec:Detection_Rate}

Although CLs are excellent tracers of AGN activity, they do not always appear in AGN spectra and a number of reasons have been proposed to explain this. In general, CL detections decrease with increasing IP and our results agree with this; we do not detect any of the high-IP CLs, [\ion{Fe}{13}] 1.0747 $\mu$m (330.8 eV) and [\ion{S}{11}] 1.9196 $\mu$m (447.1 eV). 
Stellar absorption features can also affect detection rates by attenuating the CL emission profile. Such is the case with [\ion{Ca}{8}] which falls within the CO (3-1) absorption feature. In addition, Ca and Al species can be affected by metallicity and depletion onto dust. Moreover, telluric absorption from the atmosphere always has to be contended with in ground-based observations.

In nearby low-luminosity AGN, circumnuclear stellar populations can dominate the NIR continuum and thus drown out any CL emission. Bright AGN, particularly at high redshift, can also hamper CL emission due to their strong continuum. These effects can be seen in the results of \citet{Rodriguez2011} who found that many galaxies with stars contributing $\sim$90$\%$ of the continuum do not show CL emission. They also found their CL detection rate decreased by 17$\%$ when selecting galaxies with a redshift $>$ 0.05.

Our coverage of $J$, $H$, and $K$-bands allows us to estimate the contribution of circumnuclear stellar populations. The CO(6-3) absorption at 1.62 $\mu$m can provide an estimate to the flux contribution of red giants to the $H$-band continuum \citep{Martins2010}, where the continuum level arises due to stellar and AGN contribution. For a population of GKM giants, the typical observed depth of the absorption is $\sim$20$\%$ of the continuum \citep{Schinnerer1998}. Our sample ranges from 7$\%$ - 14$\%$ (median of 11$\%$), suggesting a large contribution to the $H$-band continuum from red giants, up to 70$\%$. The weighted average of the depth of galaxies with no CL detections is 28$\%$ deeper than those with CL emission. One possibility is that a larger population of red giants near the center could be the cause of the deeper absorption, and are thus directly increasing the continuum level. Alternatively, a shallower CO absorption may be indicative of a stronger AGN contribution to the continuum, which would likely lead to stronger CL emission that is more easily detectable.

\citet{Baldassare2017a} report 0.5-7 keV X-ray luminosities for J0906+5610 and J0954+4717, from which we convert to $L_{\rm{2-10 keV}}$ using \textsc{PIMMS}\footnote{\url{https://cxc.harvard.edu/toolkit/pimms.jsp}}. This results in a $L_{\rm{2-10 keV}}$ of 2.89 $\times$ 10$^{40}$ erg s$^{-1}$ and 6.12 $\times$ 10$^{39}$ erg s$^{-1}$ for J0906+5610 and J0954+4717, respectively. Additionally, using 0.3-8 keV fluxes from \citet{Wang2016}, we calculated 2-10 keV luminosities for J0811+2328 (1.33 $\times$ 10$^{39}$ erg s$^{-1}$) and J1005+1257 (5.20 $\times$ 10$^{39}$ erg s$^{-1}$). Although J0811+2328 does have the lowest X-ray luminosity, we find no clear distinctions between galaxies with CL detections and those without. We do note, however, that these X-ray luminosities are about two orders of magnitude lower than the $L_{\rm{2-10 keV}}$ -- $L_{\rm{W2}}$ relation discussed in \citet{Secrest2015}. Indeed, all four targets have $L_{\rm{2-10 keV}}$/$L_{\rm{W2}}$ $<$ -2.1. Similar low $L_X$ have been reported in low-mass galaxies \citep{Dong2012,Simmonds2016,Cann2020}, suggesting obscuration of the X-ray source emission or that AGN in dwarf galaxies are X-ray weak compared to their MIR emission.

\subsection{Coronal Line and Outflow Kinematics} \label{subsec:Kinematics}

Photoionization is widely considered as the main excitation mechanism behind CL emission, although \citet{Rodriguez2006} found that their models more precisely matched emission line ratios from their data when shocks were included. If, however, photoionization is the principle excitation mechanism behind CLs then we expect a correlation between the FWHM and IP. Emission from a high IP line would suggest that the emitting gas is located closer to the central ionizing source and thus be deeper in the gravitational well, causing a broadening of its emission line profile. Because of their high range of IPs ($\sim$100 eV -- 500 eV), CLs are ideal in investigating the gas kinematics near the AGN. Indeed, positive correlations between line width and IP have been found in past studies \citep[eg.,][]{DeRobertis1984,DeRobertis1990,Veilleux1991} for optical high ionization lines. A positive trend has also been found between line width and critical density in these studies. Similar trends between line width and IP for NIR lines have been seen in some galaxies \citep{Rodriguez2002} but larger samples are starting to show more varied CL widths \citep[eg.,][]{Rodriguez2011,Villar_Martin2015,Cerqueira2020}, and in many cases no trends are found at all. \citet{Rodriguez2011} find a positive slope up to $\sim$300 eV, after which the slope turns negative (i.e. higher IP, lower FWHM). They attribute this to the increase in electron density when approaching the central AGN. Due to densities exceeding the critical densities of the high IP CL ions, these lines may be suppressed. Specifically, collisional de-excitation could reduce emission associated with the broader, high velocity components, thus causing us to only see the narrow emission and explaining the decrease in FWHM at high IPs.

\begin{figure*}
\centering
\epsscale{1.1}
\plotone{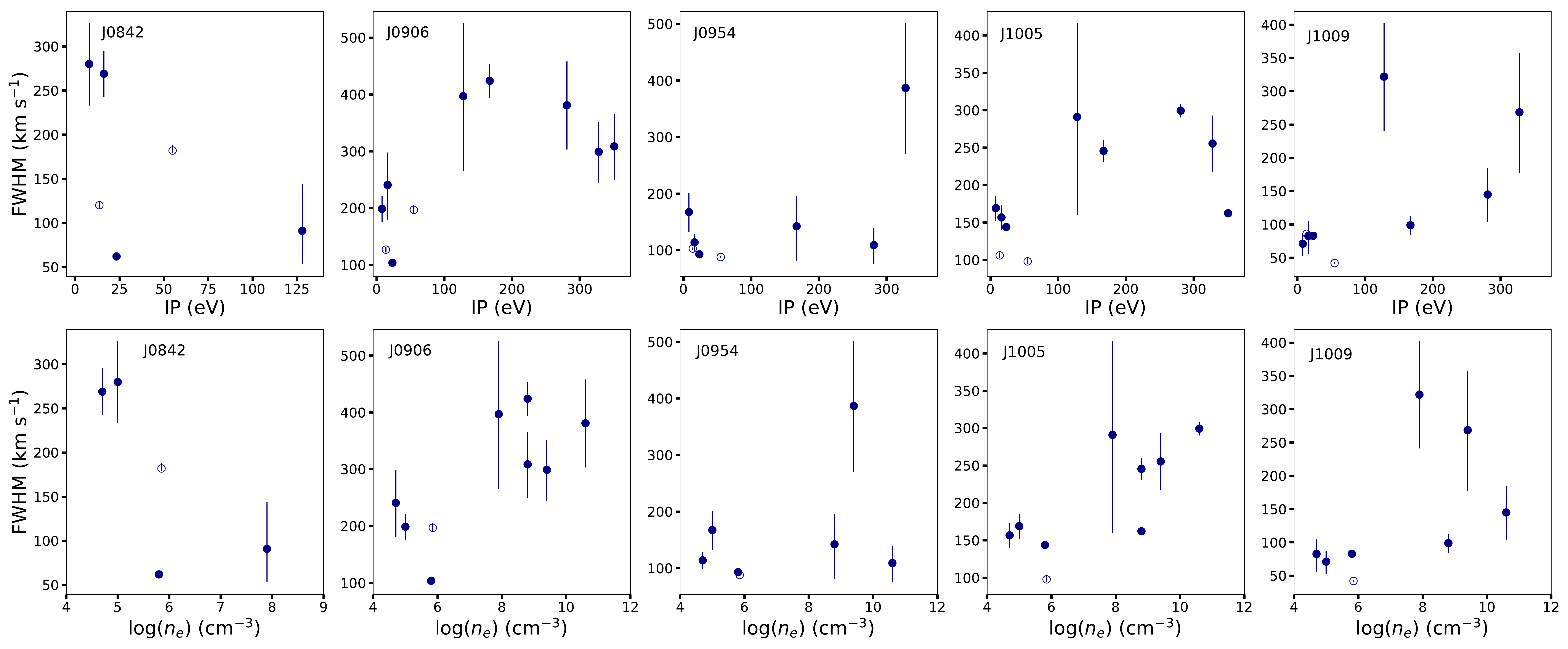}
\caption{FWHM versus IP (top panels) and critical density (bottom panels) for our sample with NIR CL detections. Low-ionization optical emission lines are plotted as open circles while all NIR emission lines are plotted as filled circles. \label{fig:IP_FWHM}}
\end{figure*}

\begin{deluxetable}{cccccc}
\caption{Emission Lines with IP and Critical Densities} 
\label{tab:Ion_Pot}
\tablehead{\colhead{Emission Line} & \colhead{IP} & \colhead{log($n_e$)} & \colhead{Emission Line} & \colhead{IP} & \colhead{log($n_e$)}\\
\colhead{ $\mu$m} & \colhead{eV} & \colhead{cm$^{-3}$} & \colhead{ $\mu$m} & \colhead{eV} & \colhead{cm$^{-3}$}}
\startdata
H$\alpha$ & 13.6 & --- & $[\rm{Fe\;II}]$1.2567 & 7.9 & 5.0\\
$[\rm{O\;III}]$0.5007 & 54.9 & 5.85 & $[\rm{Si\;X}]$1.4305 & 351.1 & 8.8\\
$[\rm{S\;III}]$0.9531 & 23.3 & 5.8 & $[\rm{Fe\;II}]$1.6435 & 16.2 & 4.7\\
$[\rm{S\;VIII}]$0.9913 & 280.9 & 10.6 & $[\rm{Si\;VI}]$1.9630 & 166.8 & 8.8\\
$[\rm{Si\;IX}]$1.2523 & 328.2 & 9.4 & $[\rm{Ca\;VIII}]$2.3214 & 127.7 & 7.9
\enddata
\end{deluxetable}

We run a similar analysis by plotting narrow-component FWHM versus IP and critical density for the CLs detected in our sample in Figure \ref{fig:IP_FWHM}. We have also included optical (open circles) and NIR low-ionization lines to draw comparisons (see Table \ref{tab:Ion_Pot}). In two galaxies, J0906+5610 and J1005+1257, the widths peak around 250 -- 300 eV and subsequently decrease, consistent with collisional de-excitation of the broader components of high IP ions. The CL emission in the other three galaxies have lower S/N so it is difficult to evaluate any trends. Aside from these low S/N measurements (as indicated by their error bars), these CLs have widths consistent with the lower IP lines. However, some uncertainty may arise from overestimating/underestimating the broad/narrow component of the line profile. For instance, the relatively small width of [\ion{Si}{6}] (166.8 eV) in J0954+4717, where we have added a secondary broad component, could be due to overestimating the broad component. Regardless, our results are consistent with the recent reports finding no clear trends between FWHM and IP. 

Critical density, defined as the density when the collision rate matches the radiative de-excitation rate, is plotted against FWHM in the bottom panels of Figure \ref{fig:IP_FWHM}. Consistent with past works \citep{DeRobertis1986,Veilleux1991b,Ferguson1997}, we find linear correlation in some galaxies, with varying degrees of slope. Regardless, we find a stronger correlation of FWHM increasing with critical density, consistent with past studies.

Table \ref{tab:outflow_vel} lists the kinematic properties of the narrow and broad (outflow) components of [\ion{O}{3}] and [\ion{Si}{6}], where the [\ion{O}{3}] values come from L20. The higher spectral resolution of the IFU data of L20 allowed for the decomposition of the emission line profile into two or three components. Generally, the C1 component traces the gas in the narrow-line region while C2 and C3 represent the broader and bluer components to the multi-component fit, and thus likely trace the outflow. For J0811+2328, only one component (C1) was fit but it is blueshifted relative to the stellar velocity and broader than the stellar velocity dispersion, and thus L20 suggest that it is likely part of the outflow. For v$_{50}$, the median velocity offset of the profile relative to systemic velocity, we take the minimum values (i.e. maximum blue offset) for [\ion{O}{3}].

The velocity of an outflow is often calculated through the use of W$\rm{_{80}}$, the width containing 80$\%$ of the flux of an emission line \citep{Harrison2014}. For a single Gaussian profile, W$\rm{_{80}}$ = 1.09$\times$FWHM. To measure W$\rm{_{80}}$ for [\ion{Si}{6}] , we use the full profile due to the relatively large uncertainties in the broad component. We can define the outflow velocity as,

\begin{equation}
\label{eq:outflow_vel}
v_{\rm{out}} = -v_{50} + \frac{\rm{W_{80}}}{2}
\end{equation}

Although C1 likely traces the narrow-line gas, L20 and \citet{Manzano2020} find cases where outflowing gas can be represented as a single-component profile. For this reason, we include $\rm{v_{out}}$ calculations for all [\ion{O}{3}] components. In addition, for [\ion{Si}{6}] $\rm{v_{out}}$, we use v$_{50}$ of the full profile if a multi-component fit was used. This is because of the large uncertainties associated with the broad component fits.

\begin{deluxetable*}{cccccccc}
\tablecaption{Outflow Properties}
\label{tab:outflow_vel}
\tablehead{\colhead{Galaxy} & \colhead{$[\rm{O\;III}]$} & \colhead{$[\rm{O\;III}]$ $\rm{v_{50}}$} & \colhead{$[\rm{Si\;VI}]$ $\rm{v_{50}}$} & \colhead{$[\rm{O\;III}]$ $\rm{W_{80}}$ ($\rm{v_{out}}$)} & \colhead{$[\rm{Si\;VI}]$ $\rm{W_{80}}$ ($\rm{v_{out}}$)} & \colhead{log(L$_{[\rm{SiVI}]}$)} & \colhead{log(L$\rm{_{AGN}}$)}\\
\colhead{ } & \colhead{Component} & \colhead{(km s$^{-1}$)} & \colhead{(km s$^{-1}$)} & \colhead{(km s$^{-1}$)} & \colhead{(km s$^{-1}$)} & \colhead{($\rm{erg\; s^{-1}}$)} & \colhead{($\rm{erg\; s^{-1}}$)}\\
\colhead{(1)} & \colhead{(2)} & \colhead{(3)} & \colhead{(4)} & \colhead{(5)} & \colhead{(6)} & \colhead{(7)} & \colhead{(8)}}
\startdata
J0100--0110 & Total & -130 & --- & 440 (350) & --- & $<$ 38.49 & 43.3\\
  & C1 & -60 & --- & 210 (165) & --- & --- &  \\
  & C2 & -240 & --- & 650 (565) & --- & --- &  \\
J0811+2328 & C1 & -60 & --- & 220 (170) & --- & $<$ 37.95 & 41.8\\
J0840+1818 & C1 & -30 & --- & 130 (95) & --- & $<$ 38.10 & 41.9\\
J0842+0319 & Total & -110 & --- & 700 (460) & --- & $<$ 38.42 & 42.7\\
  & C1 & -60 & --- & 220 (170) & --- & --- &  \\
  & C2 & -160 & --- & 750 (535) & --- & --- &  \\
J0906+5610 & Total & -50 & 90, -600 & 670 (385) & 970 (535) & 42.48 & 43.3\\
  & C1 & -50 & --- & 140 (120) & --- & --- &  \\
  & C2 & 30 & --- & 680 (310) & --- & --- &  \\
  & C3 & -150 & --- & 1250 (775) & --- & --- &  \\
J0954+4717 & Total & -10 & 17, 60 & 530 (275) & 650 (300) & 42.09 & 43.6\\
  & C1 & 0 & --- & 100 (50) & --- & --- &  \\
  & C2 & -70 & --- & 430 (285) & --- & --- &  \\
  & C3 & -80 & --- & 1100 (630) & --- & --- &  \\
J1005+1257 & Total & -60 & -70, -100 & 680 (400) & 1350 (770) & 43.04 & 43.2\\
  & C1 & -40 & --- & 120 (100) & --- & --- &  \\
  & C2 & -100 & --- & 710 (455) & --- & --- &  \\
  & C3 & -200 & --- & 1200 (800) & --- & --- &  \\
J1009+2656 & Total & -50 & -100 & 150 (125) & 108 (155) & 41.59 & 42.9\\
  & C1 & -30 & --- & 100 (80) & --- & --- &  \\
  & C2 & -60 & --- & 480 (300) & --- & --- &  
\enddata
\tablecomments{Columns: (1) Galaxy name. (2) Components of the [\ion{O}{3}] fit according to L20. In general, the C3 component traces the faster, broader outflow component while C2 traces the more narrow outflow component. C1 generally traces the gas of the narrow-line region. (3) Minimum values (i.e. maximum blue offset from the systemic velocity) of $\rm{v_{50}}$ based on [\ion{O}{3}] $\lambda$5007 measurements from L20. (4) $\rm{v_{50}}$ values for [\ion{Si}{6}]. The first value is for the narrow component, followed by the broad component. (5) Maximum $\rm{W_{80}}$ values based on [\ion{O}{3}] $\lambda$5007 measurements from L20. In parentheses are $\rm{v_{out}}$ values as defined in Equation \ref{eq:outflow_vel}. (6) $\rm{W_{80}}$ of the outflow based on the full [\ion{Si}{6}] profile. $\rm{v_{out}}$ values are in parentheses. (7) Total luminosity of [\ion{Si}{6}]. Upper limits are included for galaxies without [\ion{Si}{6}] detections (see Section \ref{subsec:coronal_detections} for details). (8) Extinction-corrected AGN luminosity as derived from [\ion{O}{3}] $\lambda$5007 in L20. Extinction values used were derived in this article.}
\end{deluxetable*}

Our data show the outflow velocities seen in [\ion{Si}{6}], measured through either W$\rm{_{80}}$ or $\rm{v_{out}}$, are generally faster than those seen in [\ion{O}{3}]. The main exception to this are the velocities seen in the C3 components. This is likely due to combining the narrow and broad components of [\ion{Si}{6}] when calculating W$\rm{_{80}}$ and v$_{50}$. Using only the broad component, we find velocities to be consistent or higher than those of C3, albeit with much higher uncertainty. This overall trend of higher velocities seen in [\ion{Si}{6}] than [\ion{O}{3}] implies a decelerating outflow, one where a high velocity wind originates near the AGN and slows down as it approaches the outer, lower ionization gas.

We also note that the objects with the fastest outflows and broadest profiles seen in [\ion{O}{3}], J0906+5610, J0954+4717, and J1005+1257, have the broadest [\ion{Si}{6}] emission. This is perhaps unsurprising since we are observing the same outflow, just at different locations.

We also list the bolometric AGN luminosities (L$\rm{_{AGN}}$) in Table \ref{tab:outflow_vel}. These values were derived from the observed [\ion{O}{3}] $\lambda$5007 flux in L20. Empirical bolometric correlation factors from \citet{Lamastra2009} were used: L$\rm{_{AGN}}$ = 87 L$_{[\rm{O\;III}]}$ for 38 $<$ log(L$_{[\rm{O\;III}]}$) $<$ 40 and L$\rm{_{AGN}}$ = 142 L$_{[\rm{O\;III}]}$ for 40 $<$ log(L$_{[\rm{O\;III}]}$) $<$ 42. To correct for extinction, we applied the extinction values listed in Table \ref{tab:extinction}. We find no strong correlation between AGN luminosity and outflow speeds but there is a trend for the more luminous AGN to have a higher number and more luminous CL detections. If we assume a simple biconical outflow starting close to the AGN \citep[for a detailed analysis, see][]{Muller2011}, highly ionized gas (i.e. CLs) could more easily be sent out to farther, possibly less obscured regions due to a fast outflow. Although it is difficult to make any firm conclusions due to the CL region being unresolved in our data, it appears that faster outflows may result in stronger and broader CL emission.

\subsection{Ionization and Origin of the Outflows} \label{subsec:Origins}

In this section, we investigate whether AGN or stellar processes (or both) are the primary source of ionization and the driving mechanism for the outflows.

\subsubsection{AGN or Stellar Ionization?} \label{subsubsec:Ionization_vs}

\begin{figure*}
\gridline{\fig{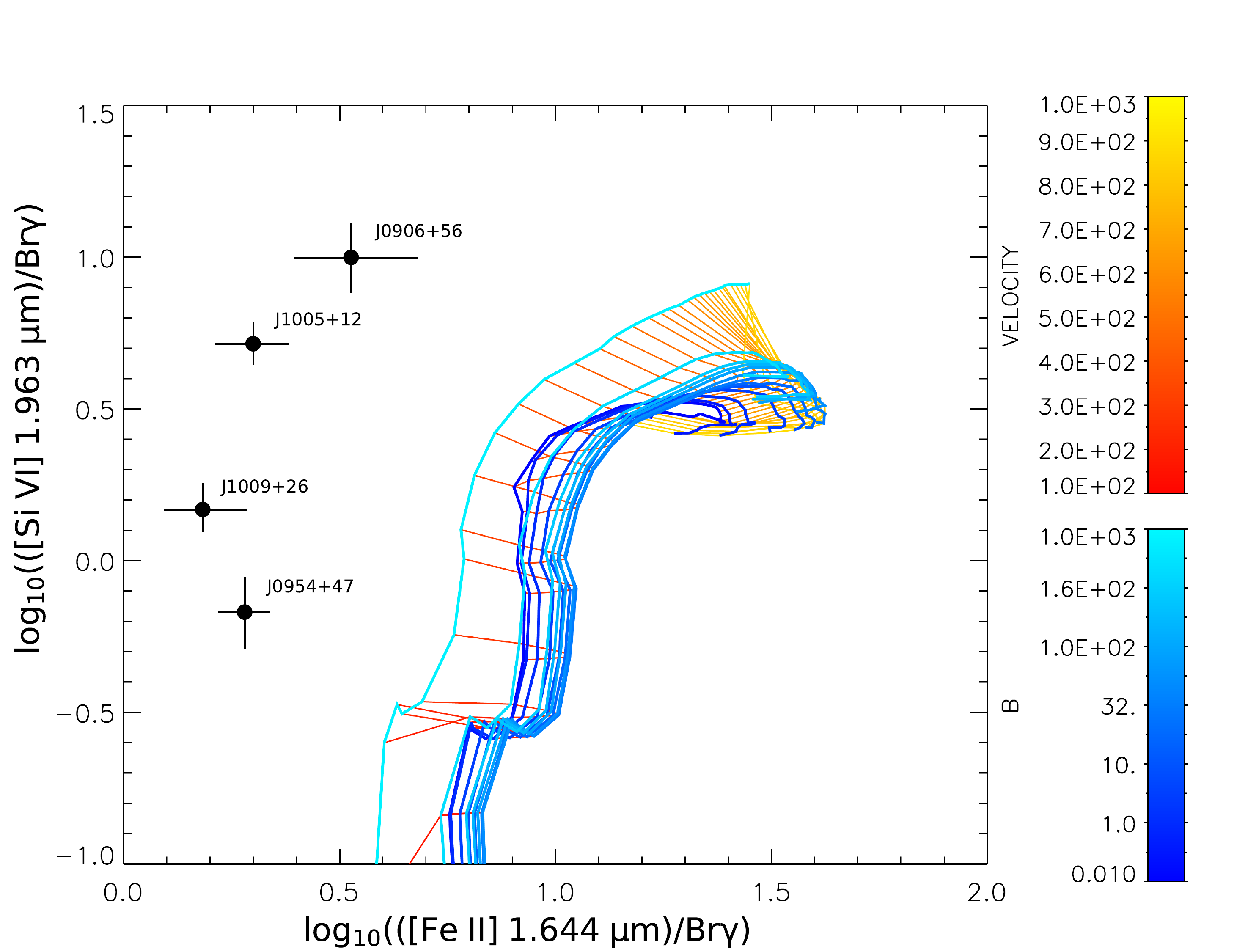}{0.5\textwidth}{(a)}
\fig{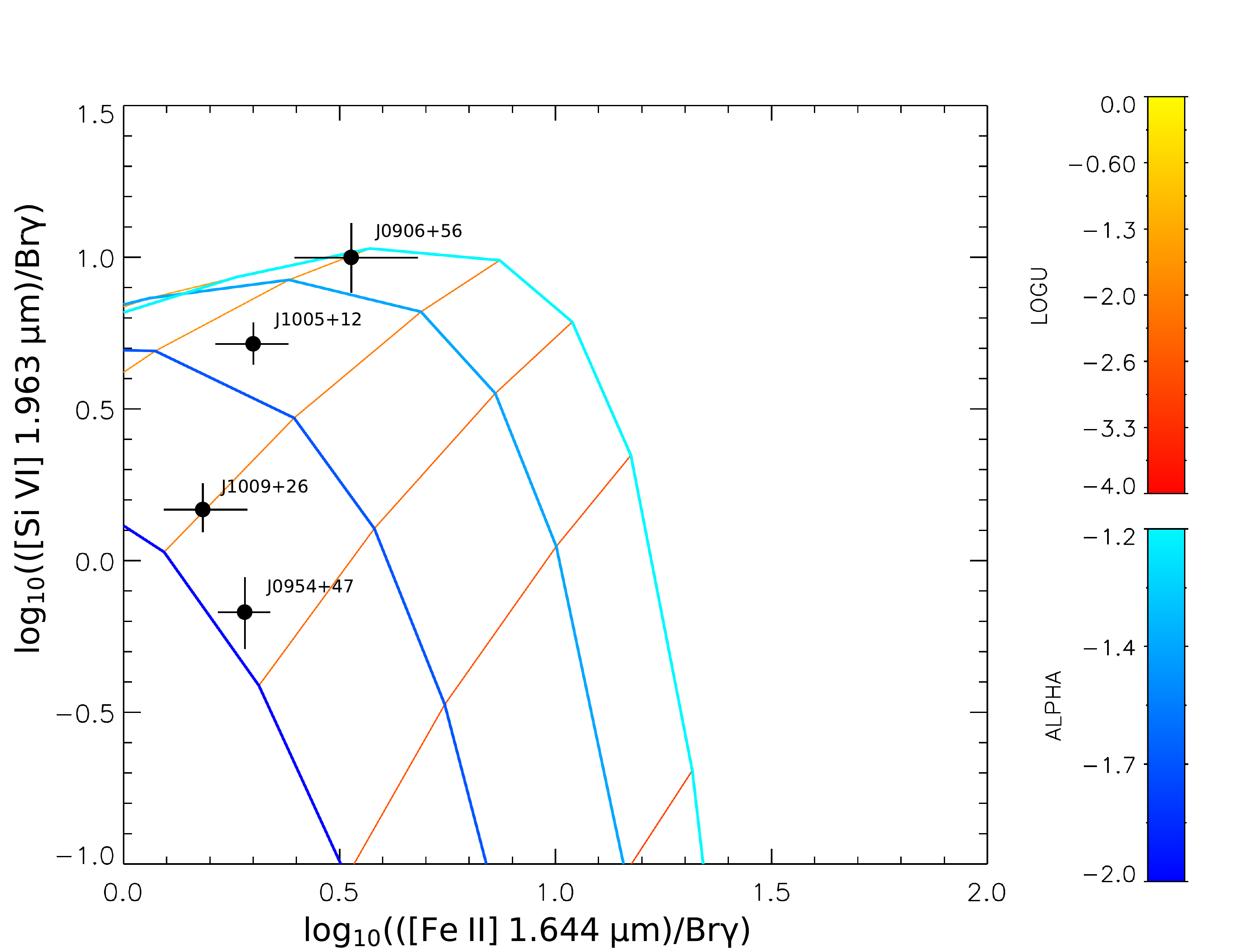}{0.5\textwidth}{(b)}}
\caption{$\rm{Log_{10}}$([\ion{Si}{6}]/$\rm{Br}\gamma)$ versus $\rm{log_{10}}$(\ion{Fe}{2}/$\rm{Br}\gamma)$ for (a) shock + precursor and (b) AGN model plots. Free parameters for the shock + precursor gridlines are the shock velocity v$\rm{_{shock}}$ and magnetic field parameter B. Free parameters for the AGN gridlines are the power law index $\alpha$ and the ionization parameter $U$. The four galaxies with detected [\ion{Si}{6}] emission are plotted as black circles with error bars. Metallicity was set to solar and a gas density of n = 1000 cm$^{-3}$ was assumed.\label{fig:shocks}}
\end{figure*}

The analysis of line ratios by L20 indicates that AGN are the primary source of ionization in our sample. However, they note that ionization from shocks, possibly originating from starburst driven-winds \citep{Sharp2010}, can not be ruled out. To investigate this further, we compare line ratios in our sample to ionization models found in the literature. \citet{Riffel2013} and \citet{Colina2015} plot [\ion{Fe}{2}] 1.64 $\mu$m/Br$\gamma$ versus $\rm{H_2}$ 2.12 $\mu$m/Br$\gamma$ to separate AGN from star-forming samples. However, the five galaxies with obtainable line ratios fall within the overlap between the AGN and SNe-dominated distributions \citep[see Figure 5 from][]{Colina2015}. We instead use the flux ratios of [\ion{Si}{6}]/Br$\gamma$ and [\ion{Fe}{2}] 1.64 $\mu$m/Br$\gamma$, where [\ion{Fe}{2}] is a sensitive indicator of shocks \citep{U2013} and Br$\gamma$ serves as an indicator of stellar activity and the ionizing radiation field. We compare these ratios to the AGN ionization models from \citet{Groves2004a,Groves2004b} and shock models from \citet{Allen2008}. We extracted these models from the \textsc{ITERA} library \citep{Groves2010} and plotted them in Figure \ref{fig:shocks}. The shock model (Figure \ref{fig:shocks}, left) takes into account both the shocked gas and precursor gas, which lies ahead of the shock front. Quick inspection of shock-only model (not plotted) shows a shift towards higher [\ion{Fe}{2}]/Br$\gamma$ ratios, farther away from our measured line ratios. We thus focus our analysis on the shock + precursor model (hereafter simply shock model) since the individual regions cannot be resolved in our data. Free parameters for the shock model include the shock velocity v$\rm{_{shock}}$ (10 -- 1000 km s$^{-1}$) and the magnetic field parameter $B/n^{1/2}$ (10$^{-4}$ -- 10 $\mu$G cm$^{3/2}$), where $B$ is the transverse magnetic field. The AGN model free parameters include the power law index $\alpha$ and the ionization parameter $U$, where $U\equiv n_{\rm{ion}}/n_{\rm{e}}$, where $n_{\rm{ion}}$ is the density of the ionizing photons and $n_{\rm{e}}$ is the electron density. The model uses a simple power law, $F_\nu \propto \nu^{\alpha}$, where 5eV $< \nu <$ 1,000eV \citep[see][for further details]{Groves2004a,Groves2004b}.

To estimate the CL gas density, \citet{Landt2015a,Landt2015b} use line ratios of the [\ion{Fe}{7}] $\lambda\lambda\lambda$3759, 5159, 6087 optical CLs. Although we do not detect all these lines in our optical data, some targets do have two [\ion{Fe}{7}] emission lines that are measurable. For these targets, we calculated an upper limit flux to the third [\ion{Fe}{7}] line and obtained a rough estimate of the gas density. These values are consistent with a density of 1000 cm$^{-3}$. In addition, L20 used the \ion{S}{2} $\lambda$6716/\ion{S}{2} $\lambda$6731 line ratio \citep[eg.,][]{Sanders2016} and obtained values $\sim$ 500 cm$^{-3}$ for most of their sample. They thus use a density of 1000 cm$^{-3}$ in their models. Therefore, in both of our models, we use a gas density of 1000 cm$^{-3}$. Lastly, metallicity was set to solar.

We use the total flux for each emission line and plot the ratios over the shock model (Figure \ref{fig:shocks}, left) and AGN model (Figure \ref{fig:shocks}, right). It is clear that the line ratios are well within the AGN model parameters (-2.0 $\lesssim$ $\alpha$ $\lesssim$ -1.2 and -3.3 $\lesssim$ log($U$) $\lesssim$ -2.0). As for the shock models, the line ratios are offset by more than 0.5 dex and all have systematically larger [\ion{Fe}{2}]/Br$\gamma$ ratios. In order to match the measured data, it appears a relatively large magnetic parameter ($> 10^3$) is required. Thus it is likely that AGN are the dominant ionizing source in our sample, though a small contribution from shocks cannot be formally ruled out.

\subsubsection{AGN or Stellar Driven Winds?} \label{subsubsec:driving_force_vs}

The matter of identifying the driving force behind the outflows is a difficult but important issue to investigate. Outflowing winds are often cited coming from either AGN or starburst activity \citep[eg.,][]{Fabian2012,Rupke2018,Veilleux2020}, though distinguishing between the two is often difficult. L20 investigated the energies associated with the outflows in our sample and found that the AGN are more than powerful enough to drive them. However, they found that typical core collapse supernovae can also provide the necessary energy output needed. 

One avenue of checking the likelihood of starburst driven-winds in our sample is to estimate the age of a stellar population. The CN absorption features at 1.1 $\mu$m have been shown to serve as an indicator of an intermediate-aged stellar population \citep{Maraston2005, Riffel2007}. This band most notably arises from carbon stars where there is an excess of carbon that is not bonded in CO molecules. To form a carbon star, a third 'dredge-up' during the thermally pulsing asymptotic giant branch (TP-AGB) is necessary and this constrains the stellar population age to be within 0.3 -- 2 Gyr. Thus CN absorption would identify a younger to intermediate aged stellar population and signify an occurrence of starburst activity. 

The absence of CN absorption, however, may not necessarily imply an absence of a young to intermediate-aged stellar population. \citet{Riffel2009} and \citet{Martins2013} report cases with no CN absorption but with known intermediate populations and vice versa. But as they point out, this is not to say that CN does not trace an intermediate population. Overlap with strong emission from \ion{He}{1} 1.08 $\mu$m and Pa$\gamma$ can partially or fully obstruct the CN absorption. Telluric absorption, particularly in galaxies with low redshift (z $<$ 0.01), may also interfere with CN detection. Even with proper correction from a telluric standard, the residuals can have substantial effects on the signal to noise on the CN features.

Visual inspection of our spectra shows no interference from strong emission lines but we do find significant telluric absorption in the region of CN in two cases (see Appendix \ref{appendix:CL_detail} for more details). For the rest, this lack of CN absorption would indicate that much of the CO absorption comes from older red giants (see Section \ref{subsec:Detection_Rate}). This is also consistent with the stellar population age estimates of at least several Gyrs modeled from optical LRIS spectra \citep{Manzano_Thesis}. This apparent deficit of young stars would suggest a lack of starburst that could drive the outflows we see, making the outflows more likely to be driven by AGN.

\section{Conclusion} \label{sec:Conclusion}

We have presented Keck NIR spectroscopy of a sample of dwarf galaxies with strong evidence of AGN activity and have shown the outflows detected are likely AGN-driven. Eight galaxies were observed with Keck NIRES, providing full wavelength coverage from 0.9 -- 2.4 $\mu$m, and one galaxy with Keck NIRSPEC, with wavelength coverage from 1.97 -- 2.39 $\mu$m. The main results are summarized below.\\

\textbullet\; NIR CLs (IP $>$ 100 eV) are detected in 5/9 (55$\%$) galaxies in our sample, consistent with detection rates found in larger mass studies. Due to their high ionization potentials, CL emission is highly indicative of AGN activity. Coupled with optical and other NIR AGN diagnostics, there is strong evidence for AGN activity in these dwarf galaxies.

\textbullet\; For the four galaxies without NIR CL emission, we suspect a strong contribution from a population of red giants that could be dominating the continuum level and hamper any weak CL emission. The deeper CO(6-3) absorption at 1.62 $\mu$m in these galaxies indicates a larger population of old stars. Alternatively, the deeper CO absorption may be indicative of weaker AGN that would result in more elusive CL emission.

\textbullet\; No clear trends are found between the widths of the CLs and their IPs. Two galaxies show a decrease in width after peaking around 250 -- 300 eV. As noted in previous works, this is likely due to collisional de-excitation caused by the high density environments near the central ionizing source. In addition, we see a positive trend between the FWHM and the critical densities in these two galaxies. The other galaxies show CL widths that are consistent with lower IP lines and their widths are not as tightly correlated with critical densities.

\textbullet\; The outflow velocities measured from [\ion{Si}{6}] emission are generally faster than those measured from [\ion{O}{3}]. This indicates the presence of a decelerating outflow. We also find that the galaxies with the highest [\ion{Si}{6}] luminosity also have the fastest outflows measured in [\ion{O}{3}].

\textbullet\; Examination of ionization models reveals that NIR emission line ratios of our sample are more consistent with AGN models than with shock + precursor models. This indicates that AGN are the main ionizing source, though a smaller contribution from shocks cannot be formally dismissed.

\textbullet\; The lack of CN absorption at 1.1 $\mu$m suggests a lack of young/intermediate (0.3 -- 2.0 Gyr) circumnuclear stars in our sample. This goes against the scenario where the outflows are produced by starburst activity, suggesting AGN as the main driving force of the outflows.

\acknowledgments

We thank the referee for their time and helpful comments on this work. We also thank Lisa Prato for her assistance with \textsc{REDSPEC} and George Becker for assisting with the NIR reduction pipeline.

We thank Dr. Percy Gomez and Dr. Sherry Yeh for supporting our Keck observations. Partial support for this project was provided by the National Science Foundation, under grant No. AST 1817233. S.V. and W. L. acknowledge partial support for this work provided by NASA through grants HST GO-15662.001A and GO-15915.001A from the Space Telescope Science Institute, which is operated by AURA, Inc., under NASA contract NAS 5-26555. We also thank Dr. Randy Campbell and the support provided by the Keck Visiting Scholars Program.

Some of the data presented herein were obtained at the W. M. Keck Observatory, which is operated as a scientific partnership among the California Institute of Technology, the University of California and the National Aeronautics and Space Administration. The Observatory was made possible by the generous financial support of the W. M. Keck Foundation.

The authors wish to recognize and acknowledge the very significant cultural role and reverence that the summit of Mauna Kea has always had within the indigenous Hawaiian community. We are most fortunate to have the opportunity to conduct observations from this mountain.

Funding for the SDSSI/II has been provided by the Alfred P. Sloan Foundation, the Participating Institutions, the National Science Foundation, the U.S. Department of Energy, the National Aeronautics and Space Administration, the Japanese Monbukagakusho, the Max Planck Society, and the Higher Education Funding Council for England. The SDSS WebSite is \url{http://www.sdss.org/}. The  SDSS is managed by the Astrophysical Research Consortium for the Participating  Institutions. The Participating Institutions are the American Museum of Natural History, Astrophysical Institute Potsdam, University of Basel, University of Cambridge, Case Western Reserve University, University of Chicago, Drexel University, Fermilab, The Institute for Advanced Study,  the Japan Participation Group, Johns Hopkins University, The Joint Institute for Nuclear Astrophysics, the Kavli Institute for Particle Astrophysics and Cosmology, the Korean Scientist Group, the Chinese Academy of Sciences (LAMOST), Los Alamos National Laboratory, the Max Planck Institute for Astronomy (MPIA), the Max Planck Institute for Astrophysics (MPA), New Mexico State University, Ohio State University, University of Pittsburgh, University of Portsmouth, Princeton University, the United States Naval Observatory, and the University of Washington.

\software{\textsc{BADASS} \citep{Sexton2020} \url{https://github.com/remingtonsexton/BADASS2}), \textsc{pPXF} \citep{Peng2002, Peng2010}, \textsc{PyRAF} (PyRAF is a product of the Space Telescope Science Institute, which is operated by AURA for NASA), \textsc{REDSPEC} (\url{https://www2.keck.hawaii.edu/inst/nirspec/redspec.html})}

\appendix

\section{Details of Coronal Line Emission}
\label{appendix:CL_detail}

In this section, we cover the NIR spectra in our sample of dwarf galaxies, with particular detail to the CL emission and important absorption bands. The following subsections are in order of increasing RA of the galaxy. In the figures showing the full wavelength coverage of our data, we provide subplots that give additional detail to key features, such as absorption features and emission lines (or lack thereof). The measured properties of every NIR emission line detected at the 3$\sigma$ level are listed in Tables \ref{tab:J0100} through \ref{tab:J1442}. The listed uncertainties come from the random error estimates during the fitting process. 

\subsection{J010005.92-011058.89} \label{subsec:J0100}

J0100-0110 is one of only two galaxies in the composite region of the BPT diagram. Based on optical emission line measurements, \citet[][hereafter MC20]{Manzano2020} report disturbed gas that is offset from the stellar rotation curve derived from a NFW dark matter density profile \citep{Navarro1996}. Indeed, L20 report strong, blueshifted [\ion{O}{3}] emission in the southern portion of J0100-0110, seen predominately in the C2 component. They suggest this could be emission from the near side of a biconical outflow.

\begin{figure*}
\centering
\epsscale{1.0}
\plotone{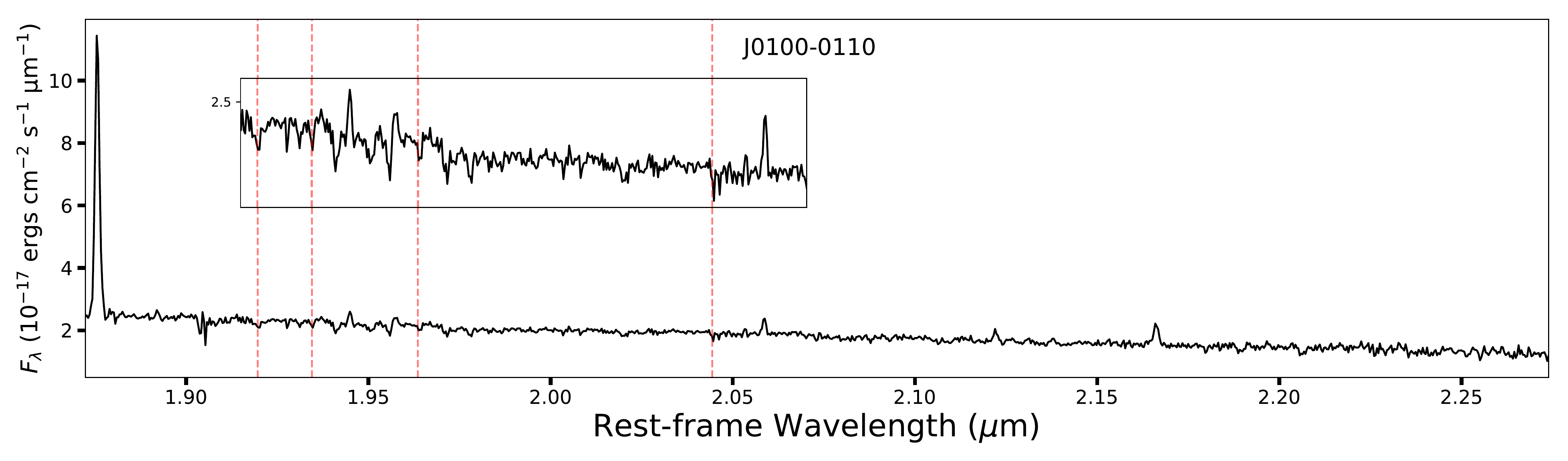}
\caption{Spectrum of J0100-0110 taken with NIRSPEC, shifted to rest-frame wavelength using the systemic redshift. Dashed red lines represent CLs that we do not detect at the 2$\sigma$ level. \label{fig:box_0100}}
\end{figure*}

The NIR spectrum of J0100-0110 is shown in Figure \ref{fig:box_0100} and line measurements are listed in Table \ref{tab:J0100}. It is the only galaxy in our sample observed with NIRSPEC (wavelength coverage from $\sim$1.97 --- 2.39 $\mu$m). Because of this, we cannot make estimates to the stellar age as we have done with the rest of the sample. The spectrum shows strong Pa$\alpha$ and Br$\gamma$; it shows no CL emission in $K$-band. No significant optical CL emission is detected in this galaxy.

\subsection{J081145.29+232825.72} \label{subsec:J0811}

\citet{Moran2014} report a lower limit to the BH mass of log($M_{\rm{BH}}/M_{\odot}$) $>$ 4.4 based on Eddington Luminosity arguments. \citet{Wang2016} report \textit{Chandra} 0.3-8 keV flux, from which we calculate a 2-10 keV lumninosity of 1.33 $\times$ 10$^{39}$ erg s$^{-1}$. MC20 report both disturbed and stratified gas in J0811+2328. Here, the gas is both offset from the stellar rotation curve, and the Balmer and forbidden lines are kinematically distinct from one another. In L20, only a single component was needed to fit the [\ion{O}{3}] profile but the larger line widths (relative to the velocity dispersion of the stellar components) and velocity offsets (up to -60 km s$^{-1}$) indicate that the outflow could be traced by this single profile.

\begin{figure*}
\centering
\epsscale{1.0}
\plotone{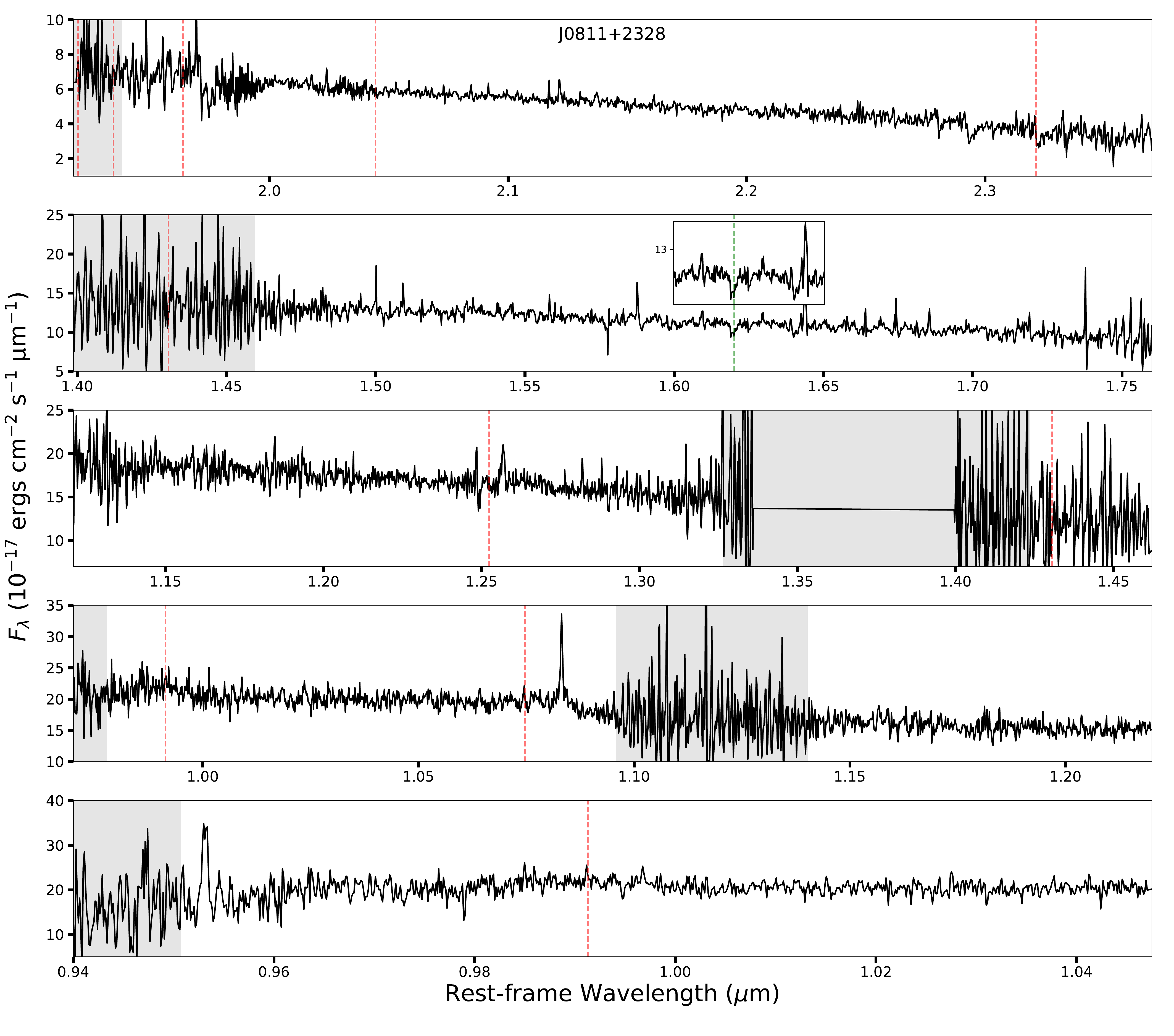}
\caption{Spectrum of J0811+2328 taken with NIRES, shifted to rest-frame wavelength using the systemic redshift. Dashed red and green lines represent CLs that we do not detect at the 2$\sigma$ level and the CO (6-3) absorption at 1.62  $\mu$m, respectively. The shaded grey regions indicate areas of significant telluric absorption. \label{fig:box_0811}}
\end{figure*}

The NIR spectrum of J0811+2328 is shown in Figure \ref{fig:box_0811} and line measurements are listed in Table \ref{tab:J0811}. We detect no NIR CL emission at the 2$\sigma$ level, though deeper observations could reveal faint [\ion{S}{8}], [\ion{Fe}{13}], [\ion{S}{9}], and [\ion{Ca}{8}]. These lines either fall on skylines or their S/N is consistent with the noise level so we do not include them in our analysis. The rest of the spectrum is relatively featureless aside from a handful of typically strong NIR lines, such as [\ion{S}{3}] 0.9531 $\mu$m and \ion{He}{1} 1.0830 $\mu$m. Unfortunately, the CN absorption at 1.1 $\mu$m falls within heavy telluric absorption so we cannot make an estimate on the contribution of a younger stellar population.

\subsection{J084025.54+181858.99} \label{subsec:J0840}

Like J0811+2328, \citet{Moran2014} also report a lower limit to the BH mass for J0840+1818, log($M_{\rm{BH}}/M_{\odot}$) $>$ 4.3. MC20 report evidence of disturbed gas and MK19 detect an outflow through a multi-component fit to [\ion{O}{3}]. Data from L20, however, show velocity offsets ranging from -30 km s$^{-1}$ to +20 km s$^{-1}$ and only a single Gaussian was used in the fit. In addition, the line widths are smaller than the velocity dispersions of the stellar component. This all suggests that the gas is rotating in the same direction as the stars and they find no clear evidence of outflows in their data.

\begin{figure*}
\centering
\epsscale{1.0}
\plotone{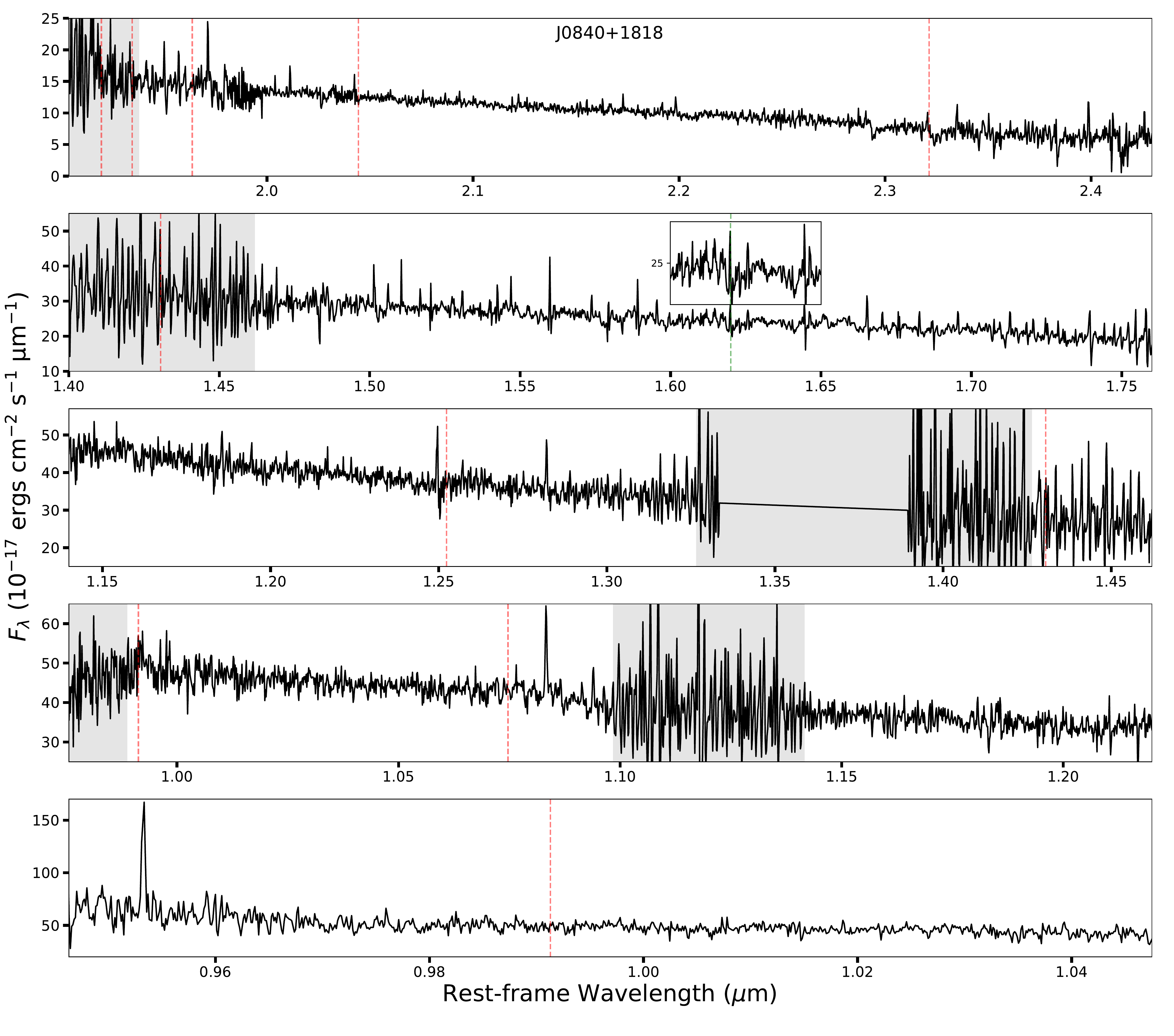}
\caption{Same as Figure \ref{fig:box_0811} but for J0840+1818. \label{fig:box_0840}}
\end{figure*}

The NIR spectrum of J0840+1818 is shown in Figure \ref{fig:box_0840} and line measurements are listed in Table \ref{tab:J0840}. We find no CLs at the 2$\sigma$ detection level, though faint [\ion{Ca}{8}] may be present. Similar to J0811+2328, we only detect a handful of emission lines and the 1.1 $\mu$m CN absorption falls within significant telluric absorption. MK19 do report the optical CL [\ion{Ne}{5}] $\lambda$3426 emission.

\subsection{J084234.50+031930.68} \label{subsec:J0842}

MK19 report broad H$\alpha$ J0842+0319, from which they calculated a BH mass of log($M_{\rm{BH}}/M_{\odot}$) = 5.84. MC20 report both disturbed and stratified gas, which is consistent with the findings of L20 where they find blueshifted gas up to -160 km s$^{-1}$.

\begin{figure*}
\centering
\epsscale{1.0}
\plotone{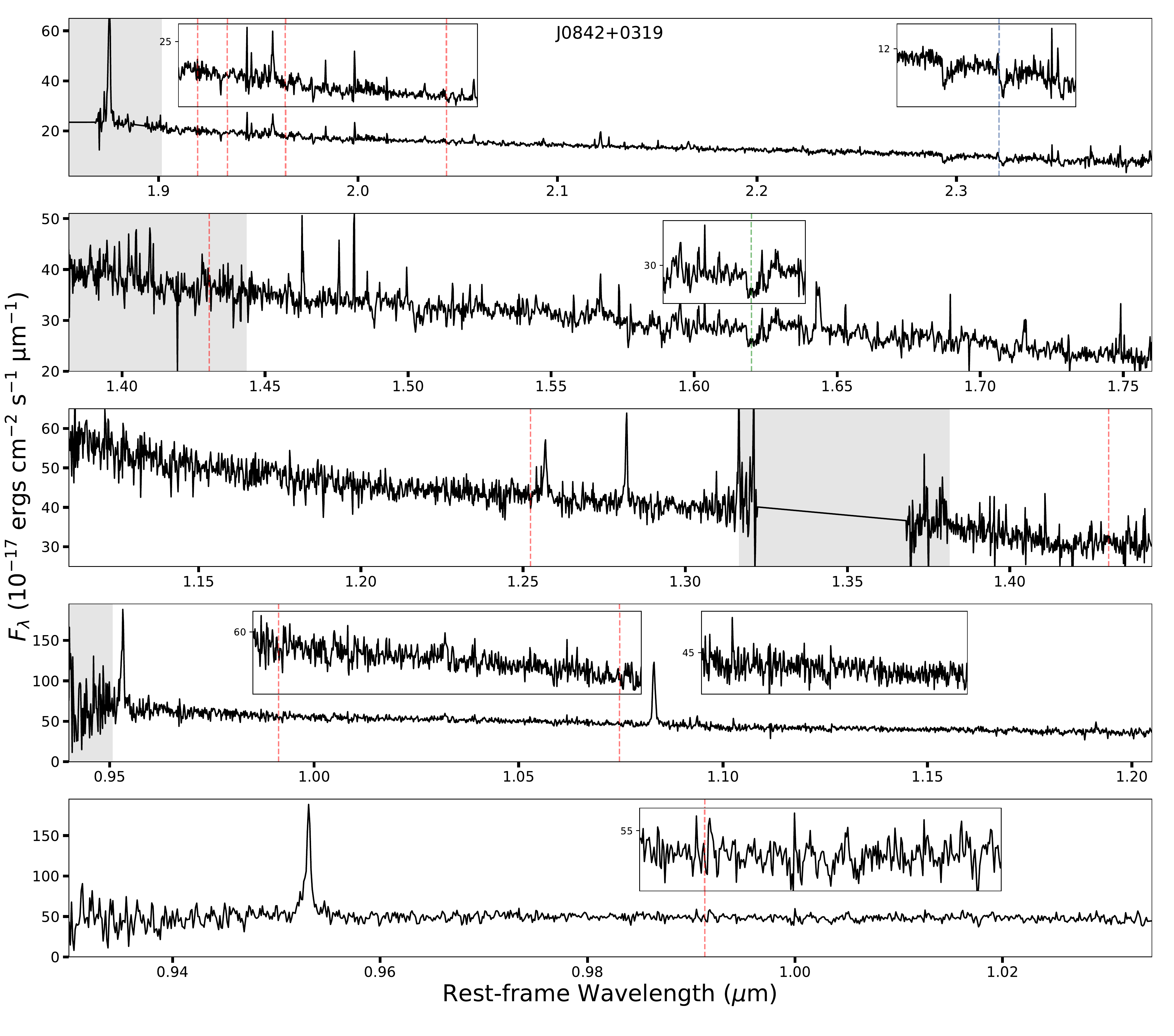}
\caption{Same as Figure \ref{fig:box_0811} but for J0842+1818. The dashed blue line indicates a CL with a 2$\sigma$ detection. \label{fig:box_0842}}
\end{figure*}

The NIR spectrum of J0842+0319 is shown in Figure \ref{fig:box_0842} and line measurements are listed in Table \ref{tab:J0842}. We detect [\ion{Ca}{8}] as the only NIR CL in this object. To provide a more accurate measurement of the flux and width, we include simultaneous fits to the CO band absorption. Of the objects that show NIR CL emission, J0842+0319 has the lowest L$_{\rm{AGN}}$ and deepest CO(6-3) band absorption feature. Deeper observations may reveal other, more faint CLs. The rest of the spectrum shows a number of hydrogen recombination, H$_2$, and \ion{He}{1} lines, and significant CN absorption is seen, suggesting an older stellar population.

\subsection{J090613.75+561015.5} \label{subsec:J0906}

\citet{Reines2015} report a BH mass of log($M_{\rm{BH}}/M_{\odot}$) = 5.40, based on virial mass estimates. \citet{Marleau2017} also report a BH mass using bolometric luminosities from WISE IR data, from which they calculate log($M_{\rm{BH}}/M_{\odot}$) = 6.93. \citet{Baldassare2017a} report bright X-ray emission, L$_{\rm{0.5-7 keV}}$ = 4.47 $\times$ 10$^{40}$ erg s$^{-1}$, in \textit{Chandra} data which is higher that what is expected from high-mass X-ray binaries. They conclude that AGN activity is the likely source of this X-ray emission. MC20 report disturbed gas and three components were used in the fits done by L20. Maximum widths of the third component reach up 1250 km s$^{-1}$, the largest seen in the sample and is indicative of a fast outflow.

We also note that the C2 component of [\ion{O}{3}] measured in L20 is redshifted 60 km s$^{-1}$ relative to the systemic velocity. One explanation of this is that they are observing the far side of the outflow. Interestingly, this redshifted component is also seen in [\ion{Si}{6}], where it is offset by $\sim$90 km s$^{-1}$. Since the two values are consistent with each other, we are both likely observing the far side of the outflow.

\begin{figure*}
\centering
\epsscale{0.97}
\plotone{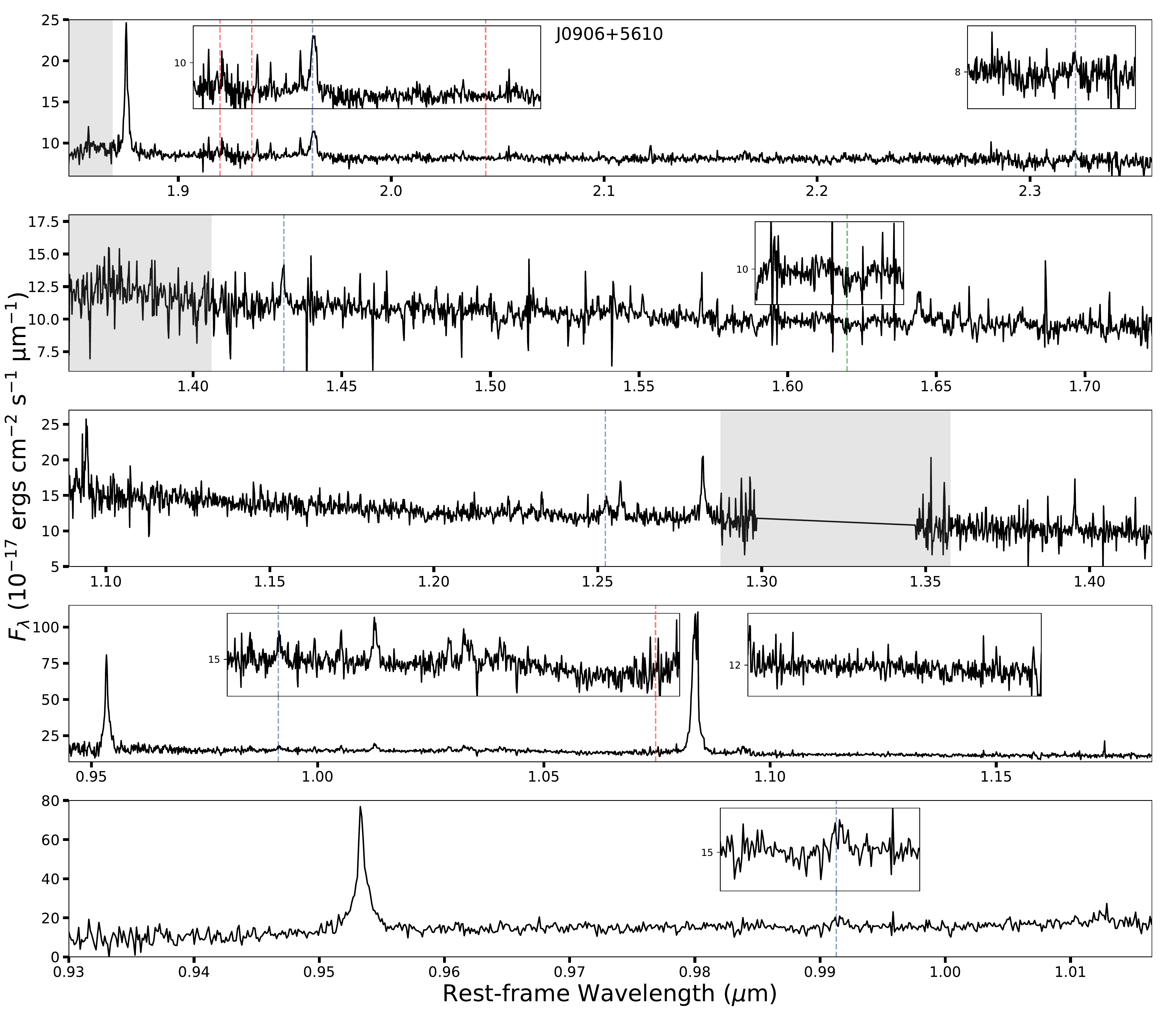}
\caption{Same as Figure \ref{fig:box_0811} but for J0906+5610. \label{fig:box_0906}}
\end{figure*}

\begin{figure*}
\centering
\epsscale{0.91}
\plotone{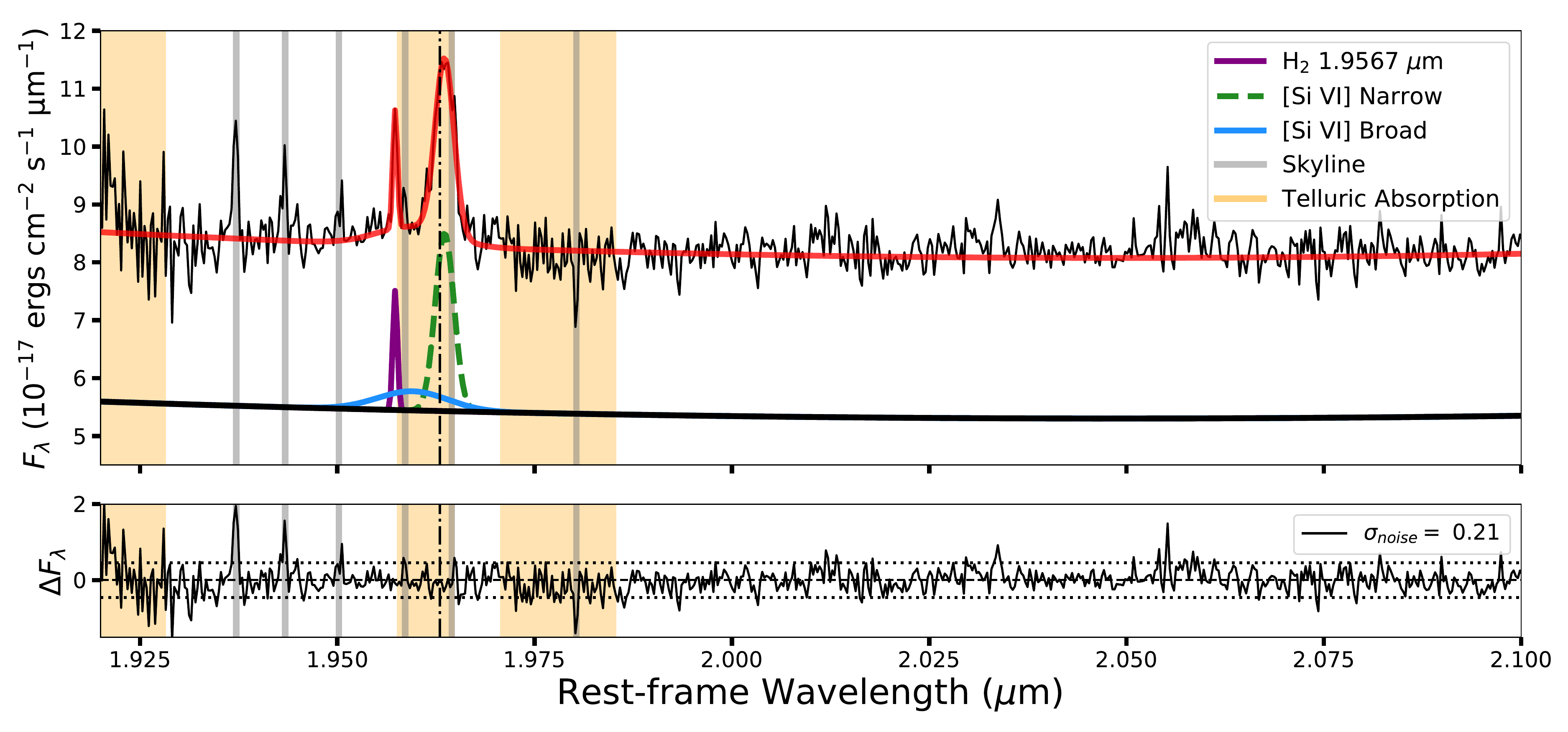}
\caption{MCMC fit to [\ion{Si}{6}]1.9630 $\mu$m and $\rm{H_2}$1.9576 $\mu$m emission for J0906. The spectrum is shifted to rest-frame wavelength using the systemic redshift. The purple line represents the fit to the $\rm{H_2}$ emission while the dotted green and blue fits represent the narrow and broad components of [\ion{Si}{6}]. The bottom panels show residuals to the fits and 1$\sigma$ noise level (represented as dotted lines). The location of skylines and regions of telluric absorption near [\ion{Si}{6}] are represented as grey and and yellow lines, respectively. Note that running a \textit{F} test on these fits suggests that a two component fit is needed. \label{fig:0906}}
\end{figure*}

The NIR spectrum of J0906+5610 is shown in Figure \ref{fig:box_0906} and line measurements are listed in Table \ref{tab:J0906}. The spectrum has an abundant number of emission lines, including strong hydrogen recombination, H$_2$, and \ion{He}{1} lines. The CL emission is also strong, with significant emission from [\ion{Si}{6}], [\ion{S}{9}], and [\ion{Si}{10}]. We also detect weaker emission from [\ion{S}{8}] and [\ion{Ca}{8}]. [\ion{Si}{6}] is best fit with two Gaussian components (see Figure \ref{fig:0906}) while only one Gaussian was necessary to fit the other CLs properly. It is noteworthy that [\ion{Si}{6}] falls within a region of telluric absorption and skylines. However, on closer inspection of the telluric standard star used, the degree of absorption is comparable to that seen redward at 1.98 $\mu$m. Thus we expect the noise level around [\ion{Si}{6}] to be at similar levels. The skylines are also similar to those blueward and the effects can be seen in the residuals of the fit. Lastly, we do not detect CN absorption in $J$-band.

Inspection of the available optical data from SDSS, LRIS, GMOS, and KCWI yields a number of CLs. These include [\ion{Ne}{5}] $\lambda$3426, [\ion{Fe}{7}] $\lambda\lambda\lambda$5721, 5159, 6087, and [\ion{Fe}{10}] $\lambda$6374. 


\subsection{J095418.16+471725.1} \label{subsec:J0954}

Both \citet{Reines2015} and \citet{Marleau2017} report a BH mass for J0954+4717. Using the methods described in Section \ref{subsec:J0906}, they report log($M_{\rm{BH}}/M_{\odot}$) equal to 5.00 and 6.46, respectively. Similar to J0906+5610, \citet{Baldassare2017a} also report strong X-ray emission, L$_{\rm{0.5-7 keV}}$ = 9.77 $\times$ 10$^{39}$ erg s$^{-1}$ that is likely coming from AGN activity. MC20 report disturbed gas and, as shown in L20, three components were needed to properly describe the profile. Both of these indicate the presence of outflows.

\begin{figure*}
\centering
\epsscale{1.0}
\plotone{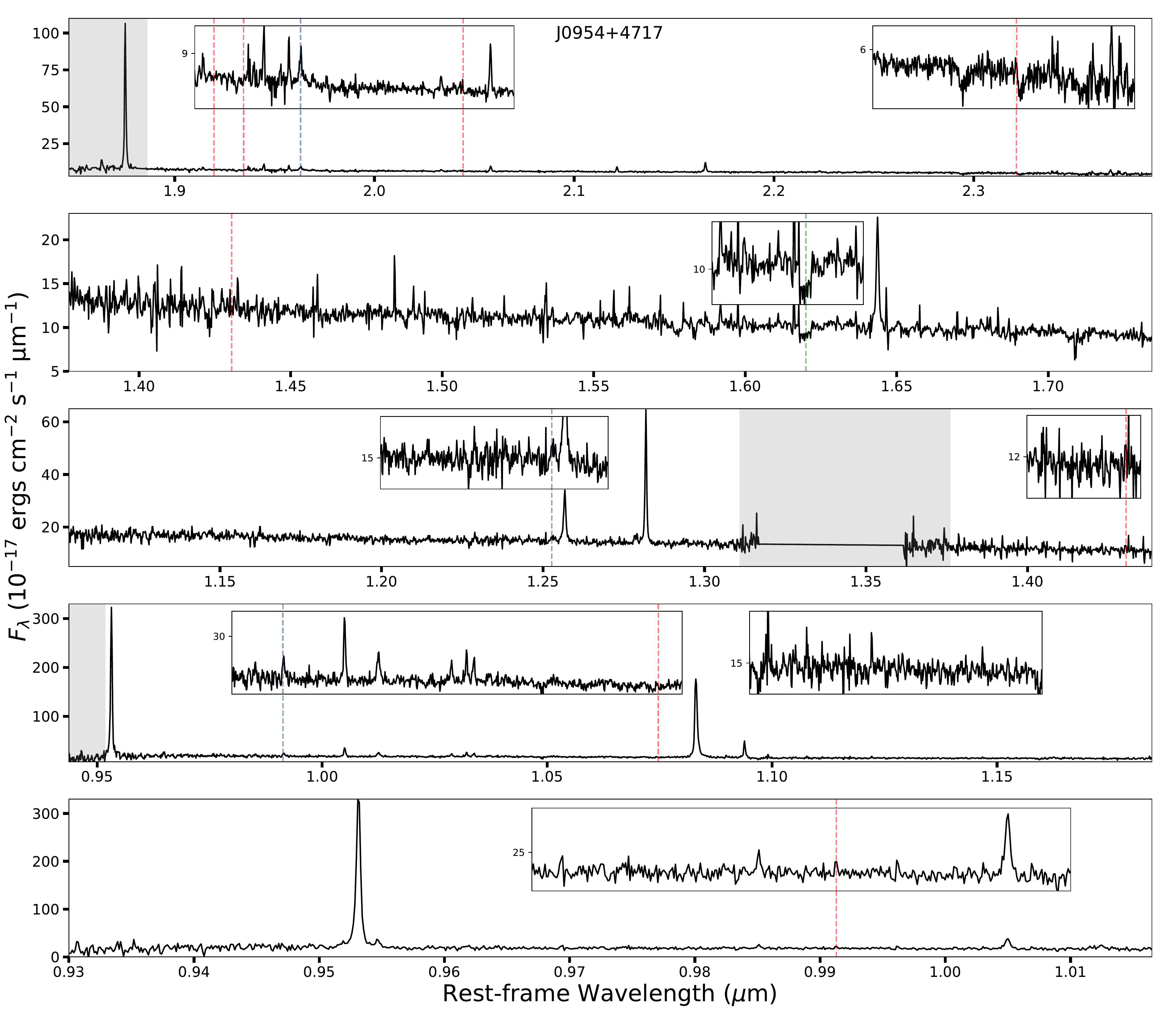}
\caption{Same as Figure \ref{fig:box_0811} but for J0954+4717.  \label{fig:box_0954}}
\end{figure*}

\begin{figure*}
\centering
\epsscale{0.95}
\plotone{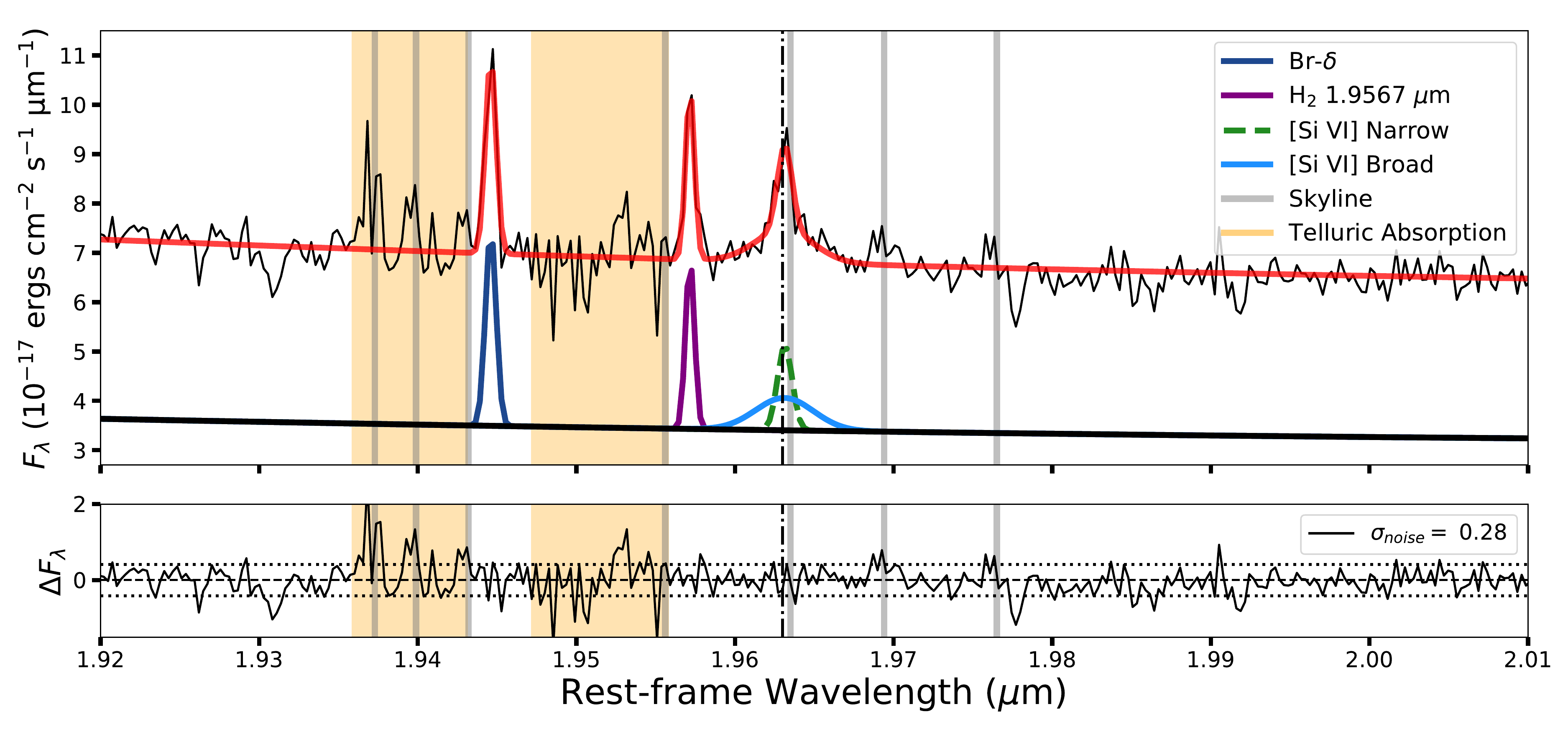}
\caption{Same as Figure \ref{fig:0906} but for J0954+4717. We additionally fit Br$\delta$, represented as a cobalt line. \label{fig:0954}}
\end{figure*}

The NIR spectrum of J0954+4717 is shown in Figure \ref{fig:box_0954} and line measurements are listed in Table \ref{tab:J0954}. Strong emission is seen in various hydrogen recombination, H$_2$, and \ion{He}{1} lines. A large number of these emission lines require a multi-component fit to properly match their profile, possibly due to the outflow. [\ion{Si}{6}], plotted in Figure \ref{fig:0954}, is the only strong CL. Inspecting the skyline immediately on top of [\ion{Si}{6}], we find it to be of similar strength to those redward. As shown in the bottom panel, the residuals at the locations of the skylines are comparable, leading us to believe our fit is representative of the emission. Evaluating the emission profile of [\ion{Si}{6}], we obtain a \textit{F}-test value of $\sim$1.5 when adding a broad component. Although this is suggestive of the presence of broad emission, we cannot say for certain whether adding a broad component is justifiable. As such, we treat this as an upper limit to the broad flux and width. 

In addition to [\ion{Si}{6}], we also detect two sulfur lines, [\ion{S}{8}] and [\ion{S}{9}]. Similar to the other galaxies in our sample, we do not detect any CN absorption in $J$-band.

Like J0906+5610, inspection of the available optical data from SDSS, LRIS, GMOS, and KCWI yields a number of CLs. These include [\ion{Ne}{5}] $\lambda\lambda$3346, 3426, [\ion{Fe}{5}] $\lambda$3839, and [\ion{Fe}{7}] $\lambda\lambda$5159, 6087.


\begin{figure*}
\centering
\epsscale{1.0}
\plotone{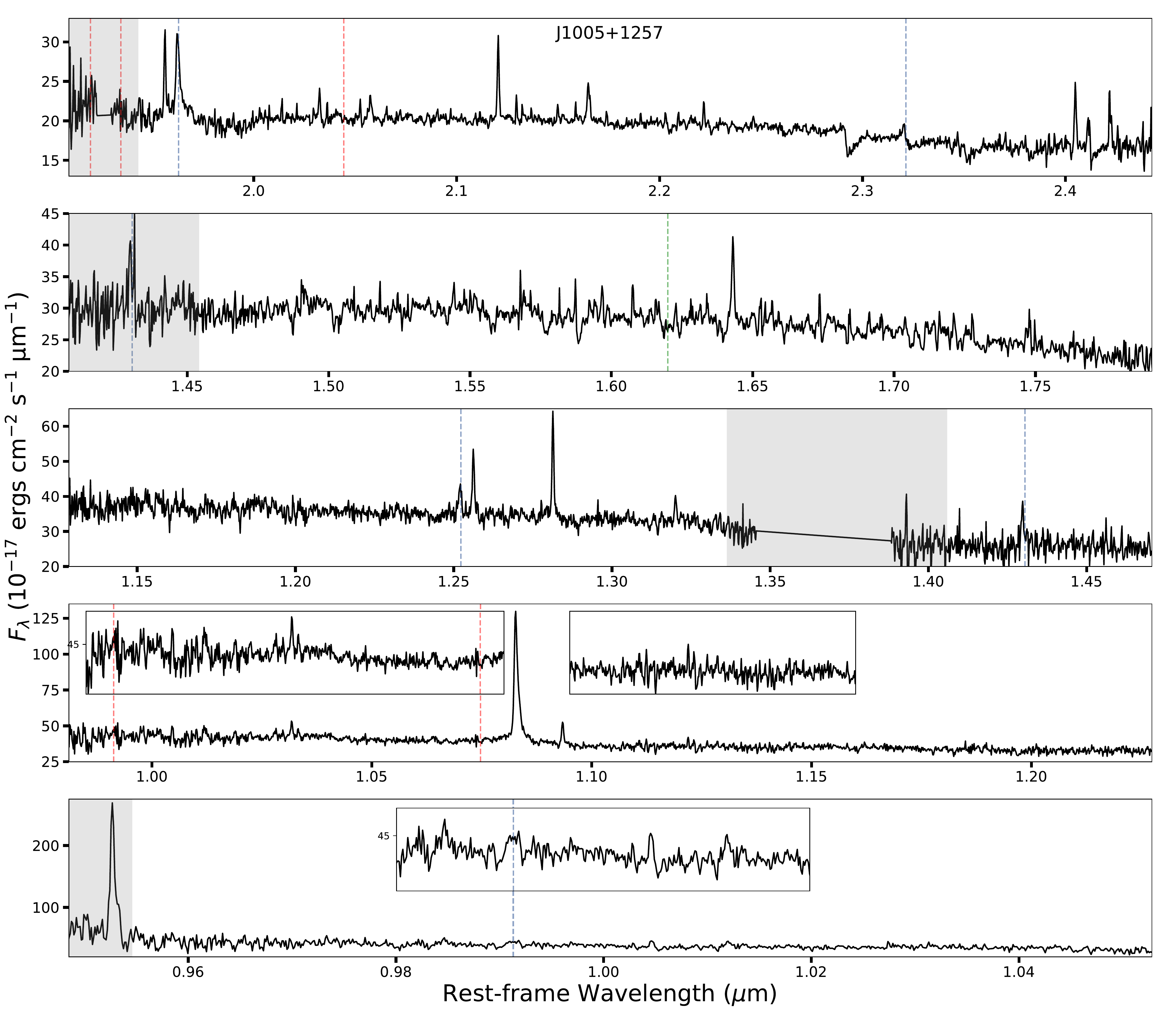}
\caption{Same as Figure \ref{fig:box_0811} but for J1005+1257.  \label{fig:box_1005}}
\end{figure*}

\begin{figure*}
\gridline{\fig{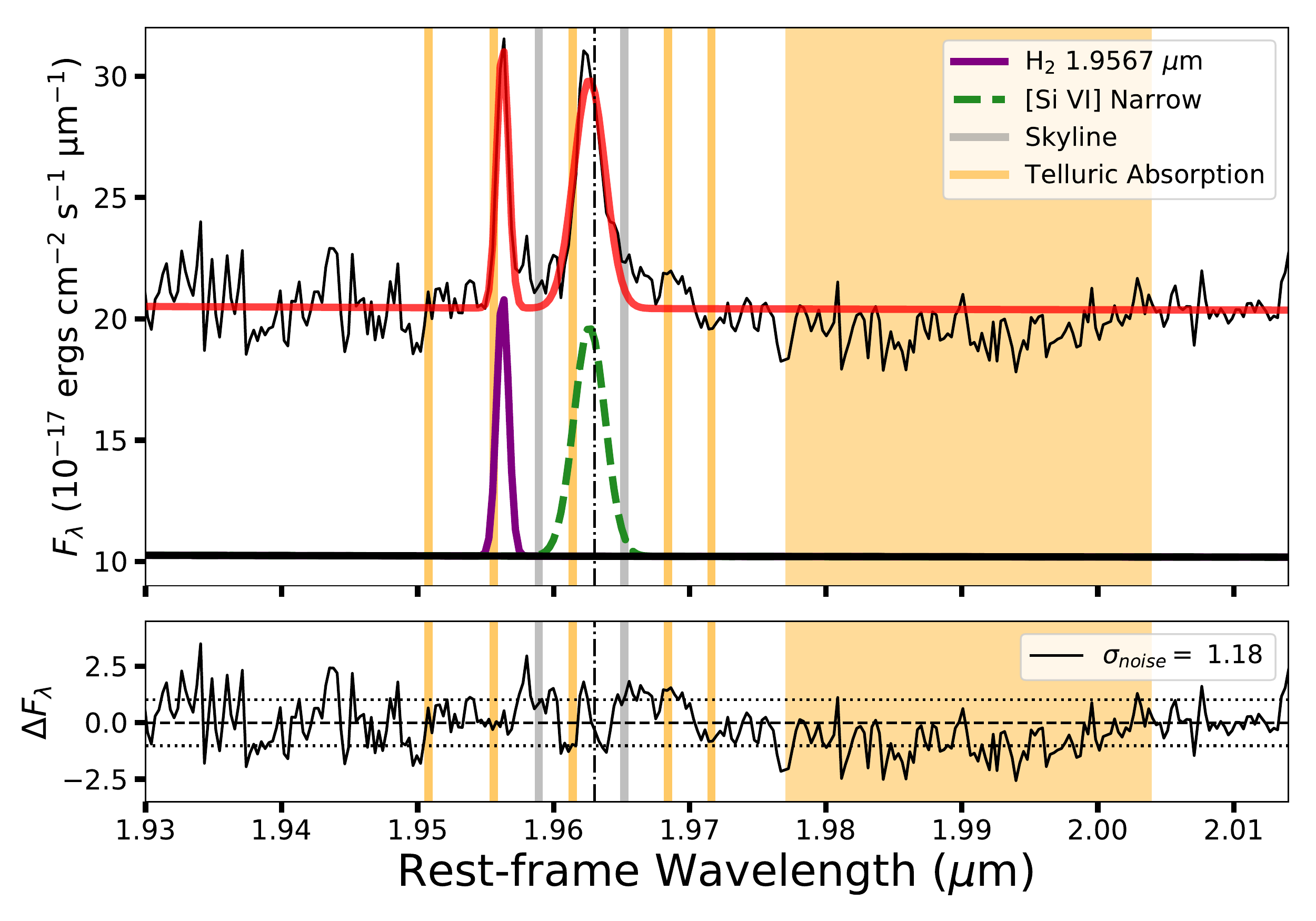}{0.5\textwidth}{(a)}
\fig{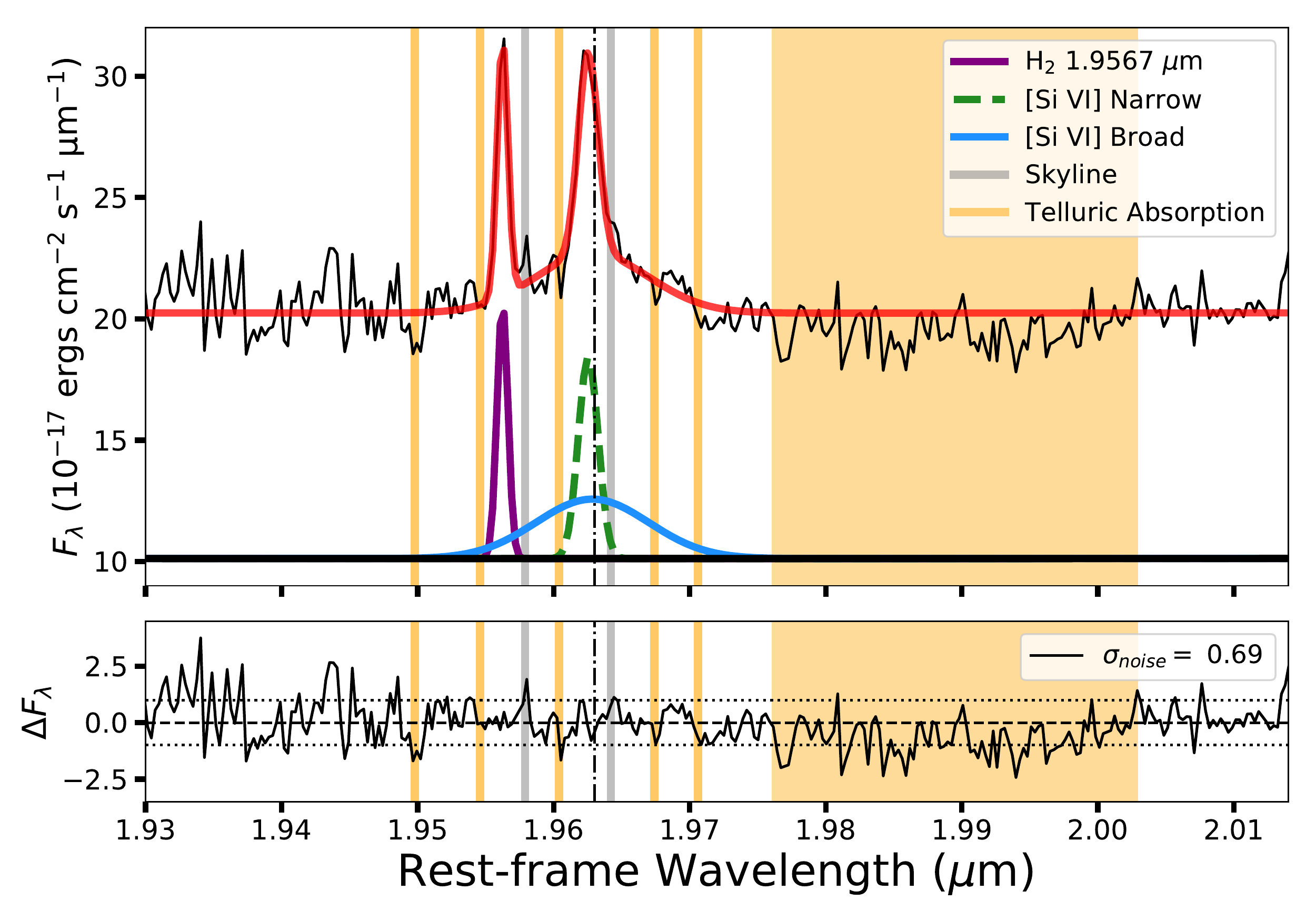}{0.5\textwidth}{(b)}}
\caption{Same as Figure \ref{fig:0906} but for J1005+1257. We plot a single component fit (left) and a double component fit (right). Note that running a \textit{F}-test on these fits suggests that a two component fit is needed.\label{fig:1005}}
\end{figure*}

\subsection{J100551.19+125740.6} \label{subsec:J1005}


\begin{figure*}
\centering
\epsscale{1.0}
\plotone{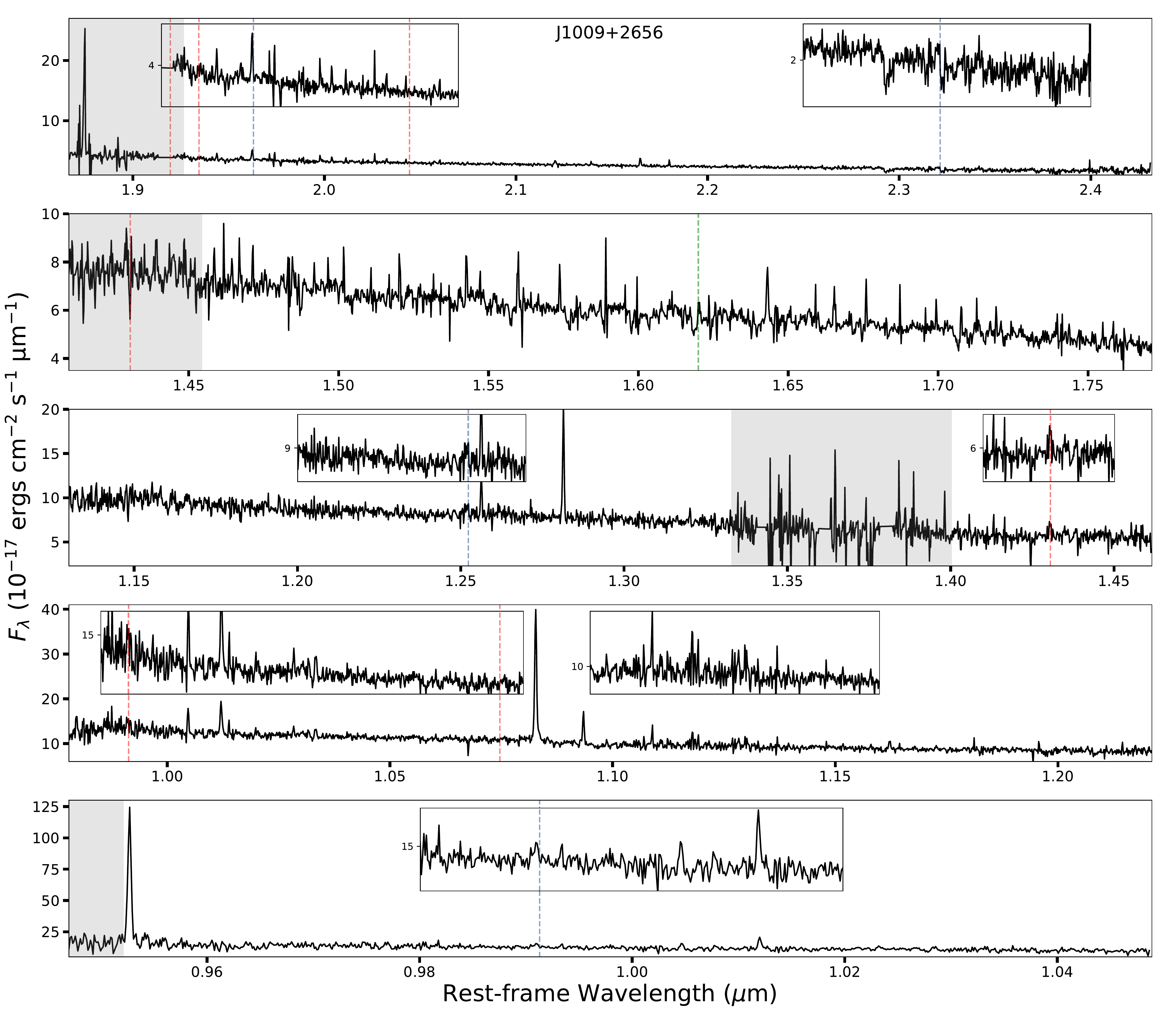}
\caption{Same as Figure \ref{fig:box_0811} but for J1009+2656.  \label{fig:box_1009}}
\end{figure*}

\begin{figure*}
\centering
\epsscale{0.95}
\plotone{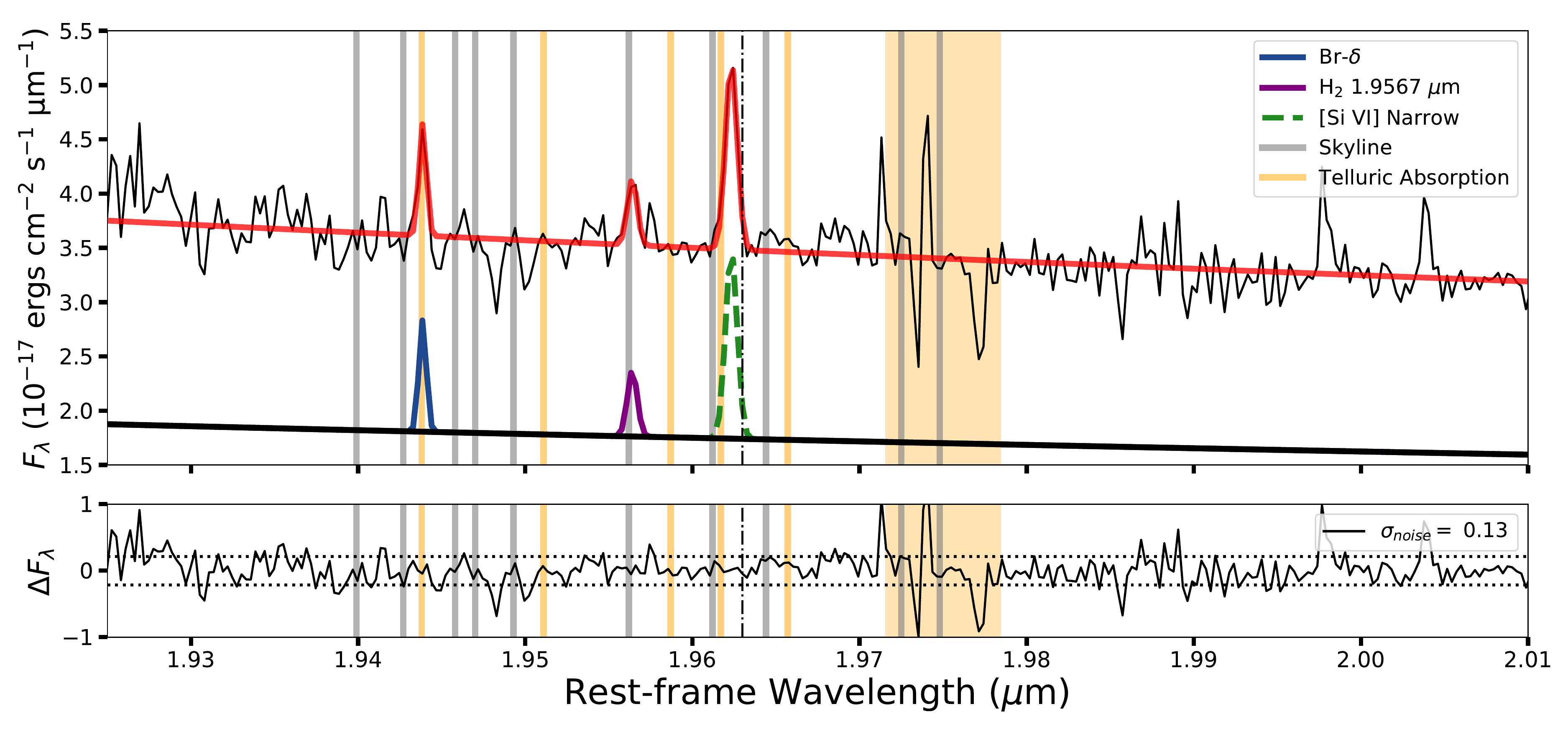}
\caption{Same as Figure \ref{fig:0954} but for J1009+2656. No broad component could be fit. \label{fig:1009}}
\end{figure*}

\citet{Marleau2017} report a BH mass of log($M_{\rm{BH}}/M_{\odot}$) = 6.47 and MC20 find disturbed gas in their optical data. Like J0811+2328, \citet{Wang2016} report a $L_{\rm{0.3-8 keV}}$ flux for J1005+1257, from which we obtain a $L_{\rm{2-10 keV}}$ of 5.20 $\times$ 10$^{39}$ erg s$^{-1}$. L20 fit three components to the [\ion{O}{3}] emission line profile, and the broadest component, C3, have widths up to 1200 km s$^{-1}$. This is the second broadest they see in the sample and suggests the presence of a fast outflow.

The NIR spectrum of J1005+1257 is shown in Figure \ref{fig:box_1005} and line measurements are listed in Table \ref{tab:J1005}. Like the other galaxies with CL emission, its spectrum shows a wealth of strong emission lines. Of particular note is [\ion{Si}{6}], which shows a strong broad component (see Figure \ref{fig:1005}, right). Unfortunately, the emission falls in a number of small telluric absorption features but they don't drastically alter the fit. The region that was most affected was the large telluric feature redward of the [\ion{Si}{6}] line. In order to not overestimate the [\ion{Si}{6}] emission, particularly the broad component, we masked out the entire telluric feature (as represented as the large yellow block) from the fit.

We also detect strong emission of [\ion{Si}{10}] and [\ion{S}{9}], and weaker emission of [\ion{S}{8}] and [\ion{Ca}{8}]. We detect [\ion{Si}{10}] in two NIRES orders and list both measurements. We select the values from order 5 since the S/N is higher and the measurements are less uncertain. Also, we do not detect any CN absorption in $J$-band.

Optical data reveal [\ion{Fe}{10}] $\lambda$6374 emission and possible [\ion{Fe}{7}] $\lambda$5721 emission. The relative abundance of NIR CL emission and lack of optical CL emission is not surprising since J1005+1257 exhibits the highest extinction of our sample (see Table \ref{tab:extinction}).

\subsection{J100935.66+265648.9} \label{subsec:J1009}

\begin{figure*}
\centering
\epsscale{0.98}
\plotone{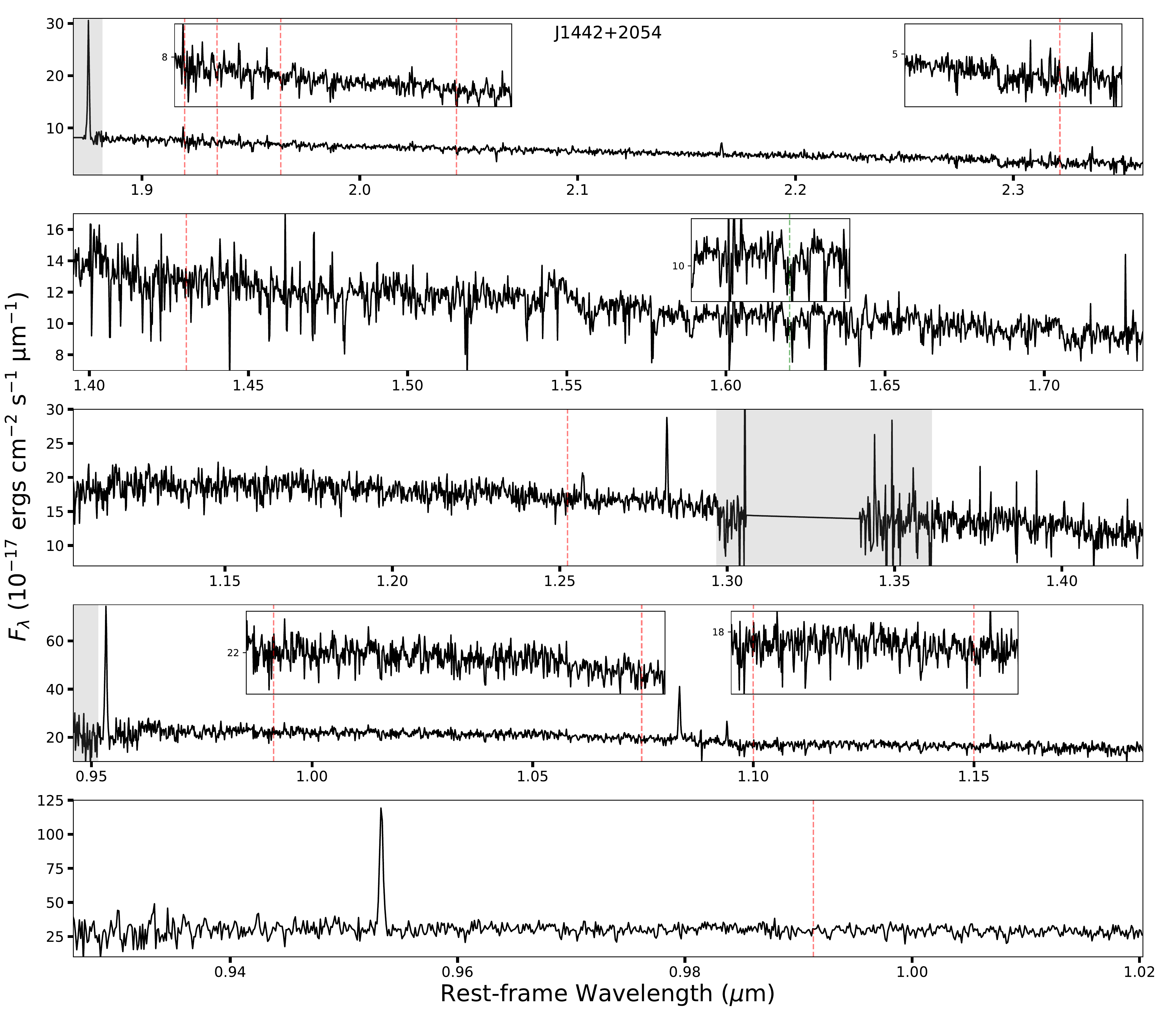}
\caption{Same as Figure \ref{fig:box_0811} but for J1442+2054.  \label{fig:box_1442}}
\end{figure*}

\citet{Moran2014} provide an upper limit to the BH mass through Eddington Luminosity arguments, log($M_{\rm{BH}}/M_{\odot}$) $>$ 5.1. In addition, MC20 report disturbed gas and a two component fit was needed to properly fit the IFU data in L20. Due to the kinematic differences and velocity offsets between the C1 and C2 components, they suggest that the C2 component could represent a tilted, biconcial outflow, similar to that of J0100-0110.

The NIR spectrum of J1009+2656 is shown in Figure \ref{fig:box_1009} and line measurements are listed in Table \ref{tab:J1009}. Only one component is needed to properly fit [\ion{Si}{6}] (see Figure \ref{fig:1009}), this being the only case in our sample. Although there appears to be a red broad component, closer inspection of the standard star reveals telluric absorption in that region which leads us to believe that it is due to a poorly corrected telluric feature. We also detect weak emission in [\ion{S}{8}], [\ion{S}{9}], and [\ion{Ca}{8}]. Lastly, we detect no CN absorption.

We also detect the optical CL [\ion{Ne}{5}] $\lambda$3426, with other possible detections including [\ion{Fe}{7}] $\lambda\lambda$5721, 6087. Deeper observations will be needed to confirm these detections.

\subsection{J144252.78+205451.67} \label{subsec:J1442}

J1442+2054 is one of two galaxies in our sample that falls in the composite region of the BPT diagram (see \citet{Manzano2019}). MC20 find both disturbed and stratified gas in their optical data. This object was not observed in L20 and no BH mass has been reported to date.

The NIR spectrum of J1442+2054 is shown in Figure \ref{fig:box_1442} and line measurements are listed in Table \ref{tab:J1442}. The NIR is relatively featureless, with only a handful of typically strong emission lines appearing. We detect no CL emission (optical nor NIR) in this galaxy. We also do not detect any CN absorption.


\startlongtable
\begin{deluxetable*}{cccc}
\tablecaption{J0100 NIR Emission Line Parameters}
\label{tab:J0100}
\tablehead{\colhead{Emission} & \colhead{FWHM} & \colhead{Flux} & \colhead{Notes}\\
\colhead{} & \colhead{(km s$^{-1}$)} & \colhead{($10^{-17}$ $\rm{erg\; cm^{-2} s^{-1}}$)} & \colhead{}}
\startdata
$\rm{Pa}\alpha$ 1.8756 $\mu$m & 62 $\pm$7 & 141 $\pm$6 & Edge of wavelength coverage\\
$\rm{Br}\gamma$ 2.1661 $\mu$m & 97 $\pm$46 & 11 $\pm$1
\enddata
\tablecomments{Only NIRSPEC (K-band) observations were done. \\ Columns: (1) Emission line. (2) FWHM of the emission line and its error after instrument correction. Uncertainties are only representative of random error in the fit. (3) Flux and error of the emission line, where uncertainties are from random errors. (4) Notes regarding the fitting. \vspace{5mm}}
\end{deluxetable*}

\startlongtable
\begin{deluxetable*}{cccc}
\tablecaption{J0811 NIR Emission Line Parameters}
\label{tab:J0811}
\tablehead{\colhead{Emission} & \colhead{FWHM} & \colhead{Flux} & \colhead{Notes}\\
\colhead{} & \colhead{(km s$^{-1}$)} & \colhead{($10^{-17}$ $\rm{erg\; cm^{-2} s^{-1}}$)} & \colhead{}}
\startdata
$[\rm{S\;III}]$ 0.9531 $\mu$m & 173 $\pm$9 & 153 $\pm$26 & Low transmission\\
$\rm{He\;I}$ 1.0830 $\mu$m Total & 137 $\pm$16 & 98 $\pm$17 & \\
$\rm{Fe\;II}$ 1.2486 $\mu$m & $19^{+42}_{-19}$ & 21 $\pm$6 & On skyline\\
$\rm{Fe\;II}$ 1.2567 $\mu$m & 265 $\pm$44 & 61 $\pm$13 & On skyline\\
$\rm{Pa}\beta$ 1.2822 $\mu$m & 51 $\pm$48 & 19 $\pm$6 & On skyline\\
$\rm{Fe\;II}$ 1.6435 $\mu$m & 110 $\pm$19 & 39 $\pm$6 & On skyline\\
$\rm{H_2}$ 2.1218 $\mu$m & $66^{+32}_{-57}$ & 11 $\pm$2 & \\
\hline \\[-1.8ex]
CO(6-3) 1.62 $\mu$m & 386 $\pm$81 & --- & Absorption Feature
\enddata
\tablecomments{Same as Table \ref{tab:J0100}. The CO(6-3) absorption feature at 1.62 $\mu$m is added at the bottom, along with the full-width half-minimum. The depth of the absorption feature measures about 13$\%$ of the continuum, suggesting roughly 65$\%$ of the $H$-band continuum comes from GMK red giants. \vspace{5mm}}
\end{deluxetable*}

\startlongtable
\begin{deluxetable*}{cccc}
\tablecaption{J0840 Emission Line Parameters}
\label{tab:J0840}
\tablehead{\colhead{Emission} & \colhead{FWHM} & \colhead{Flux} & \colhead{Notes}\\
\colhead{} & \colhead{(km s$^{-1}$)} & \colhead{($10^{-17}$ $\rm{erg\; cm^{-2} s^{-1}}$)} & \colhead{}}
\startdata
$[\rm{S\;III}]$ 0.9531 $\mu$m & 82 $\pm$2 & 410 $\pm$4 & Low transmission\\
$\rm{He\;I}$ 1.0830 $\mu$m & 89 $\pm$4 & 109 $\pm$2 & \\
$\rm{Pa}\gamma$ 1.0941 $\mu$m & $10^{+19}_{-10}$ & 32 $\pm$2 & Within telluric absorption\\
$\rm{Fe\;II}$ 1.2567 $\mu$m & 43 $\pm$31 & 28 $\pm$4 & On skyline\\
$\rm{Pa}\beta$ 1.2822 $\mu$m & $3^{+34}_{-3}$ & 57 $\pm$4 & On skyline\\
$\rm{Br}\gamma$ 2.1661 $\mu$m & 152 $\pm$140 & 11 $\pm$5 & \\
\hline \\[-1.8ex]
CO(6-3) 1.62 $\mu$m & 518 $\pm$126 & --- & Absorption Feature, on skyline
\enddata
\tablecomments{Same as Table \ref{tab:J0811}. The depth of the absorption feature measures about 14$\%$ of the continuum, suggesting roughly 70$\%$ of the $H$-band continuum comes from GMK red giants.}
\end{deluxetable*}

\startlongtable
\begin{deluxetable*}{cccc}
\tablecaption{J0842 Emission Line Parameters}
\label{tab:J0842}
\tablehead{\colhead{Emission} & \colhead{FWHM} & \colhead{Flux} & \colhead{Notes}\\
\colhead{} & \colhead{(km s$^{-1}$)} & \colhead{($10^{-17}$ $\rm{erg\; cm^{-2} s^{-1}}$)} & \colhead{}}
\startdata
$[\rm{S\;III}]$ 0.9531 $\mu$m Total & --- & 950 $\pm$14 & Order 7, \textit{F} $>$ 2\\
Narrow & 62 $\pm$3 & 341 $\pm$8 & Order 7\\
Broad & 415 $\pm$8 & 609 $\pm$11 & Order 7\\
$\rm{[S\;III}]$ 0.9531 $\mu$m Total & --- & 986 $\pm$19 & Order 6, low transmission, \textit{F} $=$ 2\\
Narrow & 62 $\pm$6 & 310 $\pm$12 & Order 6\\
Broad & 350 $\pm$6 & 677 $\pm$15 & Order 6\\
$\rm{Pa}\epsilon$ 0.9549 $\mu$m & 155 $\pm$16 & 76 $\pm$5 & Order 7, blend with $[\rm{S\;III}]\lambda$9531\\
$\rm{Pa}\epsilon$ 0.9549 $\mu$m & 171 $\pm$15 & 103 $\pm$6 & Order 6, blend with $[\rm{S\;III}]\lambda$9531\\
$\rm{Pa}\delta$ 1.0052 $\mu$m & 48 $\pm$34 & 24 $\pm$3 & Order 7\\
$\rm{Pa}\delta$ 1.0052 $\mu$m & 77 $\pm$60 & 24 $\pm$4 & Order 6\\
$\rm{[S\;II}]$ 1.0323 $\mu$m & 299 $\pm$40 & 71 $\pm$7 & \\
$\rm{[S\;II}]$ 1.0339 $\mu$m & 71 $\pm$40 & 20 $\pm$4 & \\
$\rm{He\;I}$ 1.0830 $\mu$m & 200 $\pm$6 & 618 $\pm$13 & \\
$\rm{Fe\;II}$ 1.2567 $\mu$m & 280 $\pm$46 & 159 $\pm$16 & \\
$\rm{Pa}\beta$ 1.2822 $\mu$m & 127 $\pm$12 & 152 $\pm$28 & Blue asymmetric tail\\
$\rm{Fe\;II}$ 1.6435 $\mu$m & 269 $\pm$27 & 152 $\pm$12 & On skyline\\
$\rm{Pa}\alpha$ 1.8756 $\mu$m & 192 $\pm$3 & 661 $\pm$7 & Within telluric absorption\\
$\rm{H_2}$ 1.9576 $\mu$m & 165 $\pm$21 & 84 $\pm$7 & \\
$\rm{H_2}$ 2.0338 $\mu$m & 129 $\pm$74 & 17 $\pm$5 & On skyline\\
$\rm{He\;I}$ 2.0587 $\mu$m & 64 $\pm$34 & 25 $\pm$4 & On skyline\\
$\rm{H_2}$ 2.1218 $\mu$m & 82 $\pm$13 & 53 $\pm$3 & \\
$\rm{Br}\gamma$ 2.1661 $\mu$m & 136 $\pm$29 & 30 $\pm$4 & \\
$\rm{H_2}$ 2.2235 $\mu$m & 50 $\pm$35 & 14 $\pm$5 & \\
$[\rm{Ca\;VIII}]$ 2.3214 $\mu$m & 90 $\pm$53 & 29 $\pm$13 & \\
\hline \\[-1.8ex]
CO(6-3) 1.62 $\mu$m & 723 $\pm$136 & --- & Absorption Feature
\enddata
\tablecomments{Same as Table \ref{tab:J0811}. The depth of the absorption feature measures about 13$\%$ of the continuum, suggesting roughly 65$\%$ of the $H$-band continuum comes from GMK red giants.}
\end{deluxetable*}

\startlongtable
\begin{deluxetable*}{cccc}
\tablecaption{J0906 Emission Line Parameters}
\label{tab:J0906}
\tablehead{\colhead{Emission} & \colhead{FWHM} & \colhead{Flux} & \colhead{Notes}\\
\colhead{} & \colhead{(km s$^{-1}$)} & \colhead{($10^{-17}$ $\rm{erg\; cm^{-2} s^{-1}}$)} & \colhead{}}
\startdata
$[\rm{S\;III}]$ 0.9531 $\mu$m Total & --- & 710 $\pm$6 & Order 7, \textit{F} $>$ 2\\
Narrow & 104 $\pm$3 & 168 $\pm$4 & Order 7\\
Broad & 657 $\pm$9 & 542 $\pm$5 & Order 7\\
$[\rm{S\;III}]$ 0.9531 $\mu$m Total & --- & 694 $\pm$16 & Order 6, Low transmission, \textit{F} $>$ 2\\
Narrow & 88 $\pm$8 & 163 $\pm$10 & Order 6\\
Broad & 630 $\pm$25 & 531 $\pm$12 & Order 6\\
$[\rm{S\;VIII}]$ 0.9913 $\mu$m & 277 $\pm$20 & 32 $\pm$2 & Order 7, On skyline\\
$[\rm{S\;VIII}]$ 0.9913 $\mu$m & 381 $\pm$77 & 29 $\pm$9 & Order 6, On skyline\\
$\rm{Pa}\delta$ 1.0052 $\mu$m & 92 $\pm$29 & 15 $\pm$2 & Order 6\\
$\rm{He\;II}$ 1.0126 $\mu$m & 306 $\pm$73 & 42 $\pm$8 & \\
$[\rm{S\;II}]$ 1.0289 $\mu$m & 206 $\pm$71 & 21 $\pm$5 & On skyline\\
$[\rm{S\;II}]$ 1.0323 $\mu$m & 350 $\pm$122 & 35 $\pm$9 & blend with $[\rm{S\;II}] $10339 $\mu$m \\
$[\rm{S\;II}]$ 10339 $\mu$m & 235.16 $\pm$127.78 & 16 $\pm$8 & blend with $[\rm{S\;II}]$ 1.0323 $\mu$m \\
$[\rm{N\;I}]$ 1.0400 $\mu$m & 151 $\pm$77 & 15 $\pm$5 &\\
$\rm{He\;I}$ 1.0830 $\mu$m Total & --- & 1660 $\pm$45 & On skyline, within telluric absorption, \textit{F} $>$ 2\\
Narrow & 394 $\pm$5 & 1260 $\pm$33 & \\
Broad & 1081 $\pm$81 & 401 $\pm$31 & \\
$[\rm{S\;IX}]$ 1.2523 $\mu$m & 299 $\pm$53 & 31 $\pm$4 & \\
$\rm{Fe\;II}$ 1.2567 $\mu$m & 199 $\pm$23 & 44 $\pm$3 & On skyline\\
$\rm{Pa}\beta$ 1.2822 $\mu$m & 282 $\pm$27 & 102 $\pm$6 & On skyline\\
$[\rm{Si\;X}]$ 1.4305 $\mu$m & 309 $\pm$58 & 46 $\pm$6 & \\
$\rm{Fe\;II}$ 1.6435 $\mu$m & 344 $\pm$57 & 44 $\pm$5 & On skyline\\
$\rm{Pa}\alpha$ 1.8756 $\mu$m Total & --- & 300 $\pm$19 & On skyline, \textit{F} $>$ 2\\
Narrow & 71 $\pm$30 & 66 $\pm$13 & \\
Broad & 440 $\pm$37 & 234 $\pm$14 & \\
$\rm{H_2}$ 1.9576 $\mu$m & 63 $\pm$36 & 15 $\pm$2 & On skyline\\
$[\rm{Si\;VI}]$ 1.9630 $\mu$m Total & --- & 127 $\pm$15 & On skyline, within telluric absorption, \textit{F} $>$ 2\\
Narrow & 424 $\pm$29 & 92 $\pm$7 & \\
Broad & 1683 $\pm$479 & 35 $\pm$13 & \\
$\rm{H_2}$ 2.0338 $\mu$m & 97 $\pm$44 & 7 $\pm$1 & \\
$\rm{H_2}$ 2.1218 $\mu$m & 81 $\pm$42 & 14 $\pm$3 & \\
$\rm{Br}\gamma$ 2.1661 $\mu$m & 174 $\pm$34 & 13 $\pm$3 & \\
$[\rm{Ca\;VIII}]$ 2.3214 $\mu$m & 397 $\pm$131 & 33 $\pm$8 & \\
\hline \\[-1.8ex]
CO(6-3) 1.62 $\mu$m & 508 $\pm$135 & --- & Absorption Feature, on skyline
\enddata
\tablecomments{Same as Table \ref{tab:J0811}. The depth of the absorption feature measures about 10$\%$ of the continuum, suggesting roughly 50$\%$ of the $H$-band continuum comes from GMK red giants.}
\end{deluxetable*}

\startlongtable
\begin{deluxetable*}{cccc}
\tablecaption{J0954 Emission Line Parameters}
\label{tab:J0954}
\tablehead{\colhead{Emission} & \colhead{FWHM} & \colhead{Flux} & \colhead{Notes}\\
\colhead{} & \colhead{(km s$^{-1}$)} & \colhead{($10^{-17}$ $\rm{erg\; cm^{-2} s^{-1}}$)} & \colhead{}}
\startdata
$[\rm{S\;III}]$ 0.9531 $\mu$m Total & --- & 1719 $\pm$75 & Order 7, \textit{F} $>$ 2\\
Narrow & 87 $\pm$2 & 1251 $\pm$56 & Order 7\\
Broad & 530 $\pm$36 & 468 $\pm$46 & Order 7\\
$[\rm{S\;III}]$ 0.9531 $\mu$m Total & --- & 1736 $\pm$74 & Order 6, \textit{F} $>$ 2\\
Narrow & 73 $\pm$2 & 1147 $\pm$60 & Order 6\\
Broad & 401 $\pm$21 & 590 $\pm$43 & Order 6\\
$\rm{Pa}\epsilon$ 0.9549 $\mu$m & 75 $\pm$24 & 52 $\pm$10 & Order 7, blend with $[\rm{S\;III}]\lambda$9531\\
$\rm{Pa}\epsilon$ 0.9549 $\mu$m & 129 $\pm$42 & 64 $\pm$13 & Order 6\\
$[\rm{C\;I}]$ 0.9853 $\mu$m & 75 $\pm$45 & 24 $\pm$8 & Order 7, On skyline\\
$[\rm{S\;VIII}]$ 0.9913 $\mu$m & 109 $\pm$30 & 30 $\pm$5 & Order 6\\
$\rm{Pa}\delta$ 1.0052 $\mu$m & 119 $\pm$10 & 108 $\pm$8 & Order 7, On skyline\\
$\rm{Pa}\delta$ 1.0052 $\mu$m & 102 $\pm$12 & 89 $\pm$8 & Order 6, On skyline\\
$\rm{He\;II}$ 1.0126 $\mu$m & 194 $\pm$23 & 56 $\pm$6 & \\
$[\rm{S\;II}]$ 1.0289 $\mu$m & 128 $\pm$44 & 28 $\pm$7 & \\
$[\rm{S\;II}]$ 1.0323 $\mu$m & 80 $\pm$28 & 38 $\pm$7 & \\
$[\rm{S\;II}]$ 1.0339 $\mu$m & 128 $\pm$29 & 37 $\pm$7 & \\
$\rm{He\;I}$ 1.0830 $\mu$m Total & --- & 1391 $\pm$44 & red asymmetric tail, \textit{F} $>$ 2\\
Narrow & 152 $\pm$5 & 1065 $\pm$35 & \\
Broad & 805 $\pm$61 & 325 $\pm$26 & \\
$\rm{Pa}\gamma$ 1.0941 $\mu$m & 96 $\pm$7 & 162 $\pm$10 & On skyline\\
$[\rm{S\;IX}]$ 1.2523 $\mu$m & 387 $\pm$116 & 24 $\pm$9 & On skyline\\
$\rm{Fe\;II}$ 1.2567 $\mu$m Total & --- & 197 $\pm$57 & On skyline, \textit{F} $=$ 2\\
Narrow & 125 $\pm$19 & 102 $\pm$39 & \\
Broad & 400 $\pm$124 & 95 $\pm$42 & \\
$\rm{Pa}\beta$ 1.2822 $\mu$m Total & --- & 342 $\pm$53 & On skyline, \textit{F} $>$ 2\\
Narrow & 59 $\pm$18 & 186 $\pm$40 & \\
Broad & 252 $\pm$64 & 157 $\pm$35 & \\
$\rm{Fe\;II}$ 1.6435 $\mu$m Total & --- & 150 $\pm$21 & On skyline, \textit{F} $>$ 2\\
Narrow & 114 $\pm$16 & 90 $\pm$15 & \\
Broad & 440 $\pm$115 & 61 $\pm$15 & \\
$\rm{Pa}\alpha$ 1.8756 $\mu$m Total & --- & 952 $\pm$36 & Within telluric absorption, \textit{F} $>$ 2\\
Narrow & 63 $\pm$5 & 655 $\pm$27 & \\
Broad & 309 $\pm$30 & 298 $\pm$24 & \\
$\rm{Br}\delta$ 1.9451 $\mu$m & 83 $\pm$11 & 33 $\pm$2 & Within telluric absorption\\
$\rm{H_2}$ 1.9576 $\mu$m & 64 $\pm$28 & 25 $\pm$3 & \\
$[\rm{Si\;VI}]$ 1.9630 $\mu$m Total & --- & 53 $\pm$11 & On skyline, \textit{F} $\sim$ 1.5\\
Narrow & 142 $\pm$54 & 19 $\pm$7 & \\
Broad & 687 $\pm$164 & 35 $\pm$8 & \\
$\rm{H_2}$ 2.0338 $\mu$m & 117 $\pm$29 & 11 $\pm$2 & \\
$\rm{He\;I}$ 2.0587 $\mu$m & 111 $\pm$8 & 35 $\pm$2 & \\
$\rm{H_2}$ 2.0735 $\mu$m & 105 $\pm$60 & 4 $\pm$2 & \\
$\rm{H_2}$ 2.1218 $\mu$m & 71 $\pm$12 & 28 $\pm$2 & \\
$\rm{Br}\gamma$ 2.1661 $\mu$m Total & --- & 79 $\pm$10 & \textit{F} $>$ 2\\
Narrow & 60 $\pm$19 & 45 $\pm$7 & \\
Broad & 374 $\pm$94 & 35 $\pm$7 & \\
$\rm{H_2}$ 2.2235 $\mu$m & 160 $\pm$39 & 12 $\pm$2 & \\
\hline \\[-1.8ex]
CO(6-3) 1.62 $\mu$m & 553 $\pm$63 & --- & Absorption Feature
\enddata
\tablecomments{Same as Table \ref{tab:J0811}. The depth of the absorption feature measures about 11$\%$ of the continuum, suggesting roughly 55$\%$ of the $H$-band continuum comes from GMK red giants.}
\end{deluxetable*}

\startlongtable
\begin{deluxetable*}{cccc}
\tablecaption{J1005 Emission Line Parameters}
\label{tab:J1005}
\tablehead{\colhead{Emission} & \colhead{FWHM} & \colhead{Flux} & \colhead{Notes}\\
\colhead{} & \colhead{(km s$^{-1}$)} & \colhead{($10^{-17}$ $\rm{erg\; cm^{-2} s^{-1}}$)} & \colhead{}}
\startdata
$[\rm{S\;III}]$ 0.9531 $\mu$m  & 144 $\pm$4 & 2504 $\pm$220 & Low transmission, red asymmetric tail\\
$[\rm{C\;I}]$ 0.9853 $\mu$m & 215 $\pm$18 & 134 $\pm$15 & \\
$[\rm{S\;VIII}]$ 0.9913 $\mu$m & 299 $\pm$9 & 142 $\pm$14 & On skyline\\
$\rm{Pa}\delta$ 1.0052 $\mu$m & 77 $\pm$5 & 74 $\pm$8 & Order 7\\
$\rm{He\;II}$ 1.0126 $\mu$m & 173 $\pm$53 & 105 $\pm$30 & Order 7\\
$\rm{He\;II}$ 1.0126 $\mu$m & 202 $\pm$36 & 102 $\pm$22 & Order 6\\
$[\rm{S\;II}]$ 1.0289 $\mu$m & 185 $\pm$45 & 66 $\pm$15 & \\
$[\rm{S\;II}]$ 1.0323 $\mu$m & 145 $\pm$20 & 134 $\pm$19 & On skyline\\
$[\rm{S\;II}]$ 1.0339 $\mu$m & 109 $\pm$36 & 59 $\pm$13 & \\
$\rm{He\;I}$ 1.0830 $\mu$m Total & --- & 2433 $\pm$194 & On skyline, \textit{F} $>$ 2\\
Narrow & 243 $\pm$5 & 1458 $\pm$96 & \\
Broad & 1337 $\pm$89 & 976 $\pm$117 & \\
$\rm{Pa}\gamma$ 1.0941 $\mu$m & 130 $\pm$11 & 179 $\pm$19 & \\
$[\rm{S\;IX}]$ 1.2523 $\mu$m & 256 $\pm$38 & 160 $\pm$25 & \\
$\rm{Fe\;II}$ 1.2567 $\mu$m & 169 $\pm$17 & 251 $\pm$26 & On skyline\\
$\rm{Pa}\beta$ 1.2822 $\mu$m & 133 $\pm$9 & 348 $\pm$28 & \\
$\rm{Fe\;II}$ 1.3208 $\mu$m & 143 $\pm$6 & 92 $\pm$6 & On skyline\\
$[\rm{Si\;X}]$ 1.4305 $\mu$m & 162 $\pm$4 & 164 $\pm$8 & Order 5, within telluric absorption\\
$[\rm{Si\;X}]$ 1.4305 $\mu$m & 156 $\pm$18 & 168 $\pm$20 & Order 4, within telluric absorption\\
$\rm{Fe\;II}$ 1.6435 $\mu$m & 157 $\pm$17 & 188 $\pm$17 & \\
$\rm{H_2}$ 1.9576 $\mu$m & 114 $\pm$11 & 135 $\pm$11 & Within telluric absorption\\
$[\rm{Si\;VI}]$ 1.9630 $\mu$m Total & --- & 491 $\pm$31 & On skyline, within telluric absorption, \textit{F} $>$ 2\\
Narrow & 246 $\pm$14 & 193 $\pm$16 & \\
Broad & 1376 $\pm$114 & 298 $\pm$27 & \\
$\rm{H_2}$ 2.0338 $\mu$m & 106 $\pm$35 & 43 $\pm$8 & \\
$\rm{He\;I}$ 2.0587 $\mu$m & 202 $\pm$56 & 47 $\pm$10 & \\
$\rm{H_2}$ 2.1218 $\mu$m & 121 $\pm$7 & 137 $\pm$7 & \\
$\rm{Br}\gamma$ 2.1661 $\mu$m & 208 $\pm$19 & 95 $\pm$8 & On skyline\\
$\rm{H_2}$ 2.2235 $\mu$m & 55 $\pm$35 & 28 $\pm$6 & \\
$[\rm{Ca\;VIII}]$ 2.3214 $\mu$m & 291 $\pm$129 & 48 $\pm$18 & \\
$\rm{H_2}$ 2.4066 $\mu$m & 104 $\pm$18 & 104 $\pm$12 & \\
\hline \\[-1.8ex]
CO(6-3) 1.62 $\mu$m & 778 $\pm$232 & --- & Absorption Feature
\enddata
\tablecomments{Same as Table \ref{tab:J0811}. The depth of the absorption feature measures about 7$\%$ of the continuum, suggesting roughly 35$\%$ of the $H$-band continuum comes from GMK red giants.}
\end{deluxetable*}

\startlongtable
\begin{deluxetable*}{cccc}
\tablecaption{J1009 Emission Line Parameters}
\label{tab:J1009}
\tablehead{\colhead{Emission} & \colhead{FWHM} & \colhead{Flux} & \colhead{Notes}\\
\colhead{} & \colhead{(km s$^{-1}$)} & \colhead{($10^{-17}$ $\rm{erg\; cm^{-2} s^{-1}}$)} & \colhead{}}
\startdata
$[\rm{S\;III}]$ 0.9531 $\mu$m & 83 $\pm$1 & 507 $\pm$20 & Order 7\\
$[\rm{S\;III}]$ 0.9531 $\mu$m & 128 $\pm$2 & 552 $\pm$26 & Order 6, Low transmission\\
$\rm{Pa}\epsilon$ 0.9549 $\mu$m & 125 $\pm$7 & 47 $\pm$4 & Order 7\\
$[\rm{S\;VIII}]$ 0.9913 $\mu$m & 145 $\pm$40 & 17 $\pm$4 & Order 7\\
$\rm{Pa}\delta$ 1.0052 $\mu$m & 44 $\pm$17 & 14 $\pm$2 & Order 7\\
$\rm{Pa}\delta$ 1.0052 $\mu$m & 26 $\pm$21 & 23 $\pm$3 & Order 6\\
$\rm{He\;II}$ 1.0126 $\mu$m & 67 $\pm$22 & 39 $\pm$6 & Order 7\\
$\rm{He\;II}$ 1.0126 $\mu$m & 114 $\pm$13 & 44 $\pm$4 & Order 6\\
$[\rm{S\;II}]$ 1.0339 $\mu$m & 92 $\pm$37 & 9 $\pm$3 & \\
$\rm{He\;I}$ 1.0830 $\mu$m & 116 $\pm$3 & 189 $\pm$8 & Red asymmetric tail, \textit{F} $<$ 2\\
$\rm{Pa}\gamma$ 1.0941 $\mu$m & 71 $\pm$30 & 35 $\pm$7 & \\
$\rm{He\;I}$ 1.1629 $\mu$m & 106 $\pm$45 & 22 $\pm$3 & \\
$[\rm{S\;IX}]$ 1.2523 $\mu$m & 269 $\pm$90 & 11 $\pm$3 & On skyline\\
$\rm{Fe\;II}$ 1.2567 $\mu$m & 71 $\pm$16 & 22 $\pm$2 & On skyline\\
$\rm{Pa}\beta$ 1.2822 $\mu$m & 83 $\pm$6 & 68 $\pm$2 & \\
$\rm{Fe\;II}$ 1.6435 $\mu$m & 83 $\pm$22 & 18 $\pm$2 & \\
$\rm{Pa}\alpha$ 1.8756 $\mu$m & 86 $\pm$2 & 197 $\pm$7 & Within telluric absorption\\
$\rm{Br}\delta$ 1.9451 $\mu$m & INDEF & 6 $\pm$1 & Within telluric absorption, width $<$ telescope spectral resolution\\
$\rm{H_2}$ 1.9576 $\mu$m & 76 $\pm$34 & 5 $\pm$1 & On skyline\\
$[\rm{Si\;VI}]$ 1.9630 $\mu$m & 99 $\pm$15 & 17 $\pm$1 & Adjacent to skyline\\
$\rm{H_2}$ 2.0338 $\mu$m & $48^{+26}_{-48}$ & 5 $\pm$1 & \\
$\rm{He\;I}$ 2.0587 $\mu$m & 117 $\pm$70 & 4 $\pm$1 & \\
$\rm{H_2}$ 2.1218 $\mu$m & 128 $\pm$26 & 8 $\pm$1 & On skyline\\
$\rm{Br}\gamma$ 2.1661 $\mu$m & 63 $\pm$22 & 12 $\pm$1 & \\
$[\rm{Ca\;VIII}]$ 2.3214 $\mu$m & 323 $\pm$79 & 12 $\pm$6 & \\
\hline \\[-1.8ex]
CO(6-3) 1.62 $\mu$m & 237 $\pm$39 & --- & Absorption Feature
\enddata
\tablecomments{Same as Table \ref{tab:J0811}. The depth of the absorption feature measures about 11$\%$ of the continuum, suggesting roughly 55$\%$ of the $H$-band continuum comes from GMK red giants.}
\end{deluxetable*}

\startlongtable
\begin{deluxetable*}{cccc}
\tablecaption{J1442 Emission Line Parameters}
\label{tab:J1442}
\tablehead{\colhead{Emission} & \colhead{FWHM} & \colhead{Flux} & \colhead{Notes}\\
\colhead{} & \colhead{(km s$^{-1}$)} & \colhead{($10^{-17}$ $\rm{erg\; cm^{-2} s^{-1}}$)} & \colhead{}}
\startdata
$[\rm{S\;III}]$ 0.9531 $\mu$m & 88 $\pm$2 & 435 $\pm$29 & Order 7\\
$[\rm{S\;III}]$ 0.9531 $\mu$m & 115 $\pm$2 & 315 $\pm$20 & Order 6, Low transmission\\
$\rm{He\;I}$ 1.0830 $\mu$m & 115 $\pm$9 & 138 $\pm$12 & In telluric absorption\\
$\rm{Pa}\gamma$ 1.0941 $\mu$m & 24 $\pm$2 & 36 $\pm$6 & In telluric absorption\\
$\rm{Fe\;II}$ 1.2567 $\mu$m & 134 $\pm$51 & 33 $\pm$8 & \\
$\rm{Pa}\beta$ 1.2822 $\mu$m & 86 $\pm$15 & 83 $\pm$9 & \\
$\rm{Fe\;II}$ 1.6435 $\mu$m & 137 $\pm$65 & 17 $\pm$5 & \\
$\rm{Pa}\alpha$ 1.8756 $\mu$m & 93 $\pm$4 & 203 $\pm$7 & On skyline, In telluric absorption\\
$\rm{He\;I}$ 2.0587 $\mu$m & 103 $\pm$60 & 8 $\pm$2 & \\
$\rm{Br}\gamma$ 2.1661 $\mu$m & 75 $\pm$26 & 21 $\pm$3 & \\
$\rm{H_2}$ 2.2477 $\mu$m & 115 $\pm$43 & 12 $\pm$2 & \\
\hline \\[-1.8ex]
CO(6-3) 1.62 $\mu$m & 446 $\pm$133 & --- & Absorption Feature, on skyline
\enddata
\tablecomments{Same as Table \ref{tab:J0811}. The depth of the absorption feature measures about 11$\%$ of the continuum, suggesting roughly 55$\%$ of the $H$-band continuum comes from GMK red giants.}
\end{deluxetable*}

\bibliography{mybib.bib}
\bibliographystyle{aasjournal}

\end{document}